\documentclass[ALICE,manyauthors]{cernphprep}
\usepackage{hyperref}
\usepackage[english]{babel}
\usepackage{graphicx}
\usepackage{verbatim}
\usepackage{amsfonts}
\usepackage{makeidx}
\usepackage{color}
\usepackage{amsmath}
\usepackage{lineno}
\usepackage{epstopdf}
\usepackage{hyperref}
\usepackage{cite}
\newcommand{\sqrts}{\sqrt{s}}

\newcommand{\av}[1]{\left\langle #1 \right\rangle}

\newcommand{\TeV}{\mathrm{TeV}}

\newcommand{\mev}{\mathrm{MeV}}
\newcommand{\gev}{\mathrm{GeV}}
\newcommand{\tev}{\mathrm{TeV}}

\newcommand{\cm}{\mathrm{cm}}

\newcommand{\mum}{\mathrm{\mu m}}

\newcommand{\mb}{\mathrm{mb}}
\newcommand{\mub}{\mathrm{\mu b}}

\newcommand{\pt}{p_{\rm t}}

\newcommand{\DtoKpi}{{\rm D^0\to K^-\pi^+}}
\newcommand{\DtoKpipi}{{\rm D^+\to K^-\pi^+\pi^+}}
\newcommand{\DstartoDpi}{{\rm D^{*+}\to D^0\pi^+}}
\newcommand{\Dzero}{{\rm D^0}}
\newcommand{\Dstar}{{\rm D^{*+}}}
\newcommand{\Dplus}{{\rm D^+}}
\newcommand{\nbinv}{{\rm nb^{-1}}}

\newcommand{\dEdx}{{\rm d}E/{\rm d}x}

\begin{document}

\begin{titlepage}

\PHnumber{2011-181}                 
\PHdate{5 January 2012}              
\title{Measurement of charm production at central rapidity \\in proton--proton collisions at $\sqrts = 7~\TeV$}

\Collaboration{ALICE Collaboration%
         \thanks{See Appendix~\ref{app:collab} for the list of collaboration 
                      members}}
\ShortAuthor{ALICE Collaboration}
\ShortTitle{Charm production at central rapidity in proton--proton collisions at $\sqrts = 7~\TeV$}

\begin{abstract}
The $\pt$-differential inclusive 
production cross sections of the prompt charmed mesons $\Dzero$, $\Dplus$,
and $\Dstar$ in the rapidity range $|y|<0.5$ were measured in proton--proton collisions at 
$\sqrt{s}=7~\tev$ at the LHC using the \mbox{ALICE} detector. Reconstructing the decays $\DtoKpi$, $\DtoKpipi$, $\DstartoDpi$, and their charge conjugates, about 8,400 $\Dzero$, 2,900 $\Dplus$, and 2,600 $\Dstar$ mesons with $1<\pt<24~\gev/c$ were counted, after selection cuts, in a data sample of 
$3.14 \times 10^8$ events 
collected with a minimum-bias trigger (integrated luminosity $L_{\rm int}=5~\nbinv$). The results are described within uncertainties by predictions based on perturbative QCD.
\end{abstract}

\end{titlepage}
\setcounter{page}{2}

\section{Introduction}
\label{sec:intro}
The study of the production of hadrons containing heavy quarks, i.e. charm and beauty, 
in proton--proton (pp) collisions at LHC energies provides a way to test, in a new energy domain, calculations of quantum chromodynamics (QCD) processes based on the factorization approach.
In this scheme, cross sections are computed as a convolution of three terms:
the parton distribution functions of the incoming protons, the partonic hard scattering 
cross section
calculated as a perturbative series in the strong interaction coupling constant,
and the fragmentation function, parametrizing the relative production yield and momentum distribution for a charm quark hadronizing to a particular species of D meson.
Recent implementations of such calculations, at the perturbation level
of next-to-leading order 
or at fixed order with next-to-leading-log resummation (FONLL)~\cite{fonll} describe well the beauty production cross section measured 
 in $\rm p\overline p$ collisions at 
$\sqrt{s}=1.96~\tev$ at the FNAL Tevatron collider~\cite{cdfB,fonllBcdf,gmvfnsBcdf}
and in pp  collisions at $\sqrt{s}=7~\tev$ 
at the CERN Large Hadron Collider (LHC)~\cite{lhcbBeauty,cmsJpsi}. The production cross section of charmed hadrons (D mesons) at Tevatron is reproduced within
the theoretical uncertainties of the calculations as well~\cite{charmcdf,fonllDcdf,gmvfnsDcdf}. 
However, the comparison suggests that charm production is underestimated 
by the results obtained with the central values of the calculation parameters,
as observed also in pp collisions at the BNL RHIC collider at the lower energy of
$\sqrt{s}=200~\gev$~\cite{phenixelepp,starelepp}. In this context, it is particularly interesting to perform the comparison for charm production at the LHC energy, which is
more than three times higher than at the Tevatron.
Furthermore, at LHC energies, the measurement of charm production in the low transverse 
momentum ($\pt$) region probes the parton distribution functions of the proton 
at small values of parton fractional momentum $x$ and squared momentum transfer $Q^2$.
For illustration, using a simplified $2\to2$ kinematics at leading order, c quarks 
($m_{\rm c}\approx 1.5~\gev/c^2$) with $\pt\sim 2~\gev/c$ and rapidity $y\sim 0$ 
probe the parton distribution functions at $x\sim 7\times 10^{-4}$ and $Q^2\sim (5~\gev)^2$, where the gluon component is dominant.
In this kinematic regime, the gluon distribution may reach the level of saturation, 
leading to a measurable departure of the observed cross sections from the 
expectations based on the factorization approach (see e.g.~\cite{heralhc}).

We report on the measurement of the production cross section of the prompt charmed mesons
$\Dzero$, $\Dplus$, and $\Dstar$, 
in pp collisions at $\sqrt{s}=7~\tev$, reconstructed in the range 
$1<\pt<24$~(16 for $\Dzero$)~$\gev/c$ and $|y|< 0.5$
with the 
ALICE experimental apparatus~\cite{aliceJINST}, using data collected in 2010. 
The detector layout and the data sample 
used for the measurement are
described in section~\ref{sec:detector}.
The D meson reconstruction procedure, the raw yield extraction, and the corrections applied to obtain the production cross sections, including the estimation of the prompt D meson fraction,
 are presented in sections~\ref{sec:decay} and~\ref{sec:crosssections}. Finally, the resulting $\pt$-differential cross sections are reported in 
section~\ref{sec:results} and compared to QCD predictions.

\section{Detector layout and data collection}
\label{sec:detector}
$\Dzero$, $\Dplus$, and $\Dstar$ mesons, and their charge conjugates, 
are reconstructed from their
decays into charged hadrons in the central rapidity region 
utilizing the tracking and particle identification capabilities of the 
ALICE central barrel detectors.

The ALICE apparatus is described in detail in~\cite{aliceJINST}. 
It consists of a central barrel covering the pseudorapidity interval
$|\eta|<0.9$, a forward muon spectrometer, and a set of small detectors in the 
forward regions for trigger and event characterization purposes.
Only the detector features that are relevant for the D meson analysis are 
presented here.
The ALICE global coordinate system is right-handed, with the origin 
coinciding with the geometrical centre of the central barrel,
the $z$ axis directed along the beam line,  the $x$ axis in the LHC 
(horizontal) plane,
pointing towards the centre of the accelerator, and the $y$ axis pointing 
upward.
The central barrel detectors are positioned within a 
large solenoid magnet, with a field ${\rm B}=0.5~\mathrm{T}$, parallel to 
the beam line.
Data collected with both magnet polarities were used for this analysis.

The innermost detector of the ALICE central barrel is the Inner Tracking 
System (ITS) which is made of six cylindrical layers of silicon detectors, 
with radial distance to the beam-line between 3.9~cm and 43.0~cm.
The two innermost layers, with average radii of 3.9~cm (about 1~cm from the 
beam vacuum 
tube) and 7.6~cm, are equipped with Silicon Pixel Detectors (SPD), 
comprising 9.8$\times$10$^6$ pixels of size
 $50\,(r\phi)\times 425\,(z)~\mum^2$, with
intrinsic spatial resolution of
 $12\,(r\phi)\times 100\,(z)~\mum^2$.
The signals of the 1,200 SPD readout chips provide a fast trigger through
a programmable logic.
The two intermediate layers, at radii of 15.0 and 23.9 cm, are made of Silicon 
Drift Detectors (SDD). 
They allow one to measure the hit position along $z$ with resolution better 
than 30~$\mum$ from the centroid of signals collected on the anodes, and 
to determine the $r\phi$ coordinate from the drift 
time with a resolution that depends on the level of calibration, as discussed
below. The two outermost layers are made of
Silicon Strip Detectors (SSD), 
located at radii of 38.0 and 43.0~cm, consisting of double-sided 
silicon strip sensor modules, with an intrinsic spatial resolution 
of $20\,(r\phi)\times 830\,(z)~\mum^2$.
The total material budget of the ITS is on average 7.7\% of radiation length 
for tracks crossing the ITS perpendicularly to the detector 
surfaces ($\eta=0$)~\cite{aliceJINST,ITSalign}.
These features enable measurement of the track impact parameter (i.e. the 
distance of closest approach of the track to the primary interaction vertex) in the 
bending plane ($r\phi$) with a resolution better than 75~$\mu$m for 
transverse momenta $\pt> 1$ GeV/$c$ (see section~\ref{sec:reconstruction}).

The ITS modules were aligned using survey information, cosmic-ray 
tracks, and pp data, with the methods described in~\cite{ITSalign}.
For the residual misalignment along the $r\phi$ coordinate, a
r.m.s. of about 8~$\mum$ for SPD and 15~$\mum$ for 
SSD modules was estimated~\cite{ITSalign,RossiVertex}. 
For SDD, with the current calibration level, the space point resolution along 
$r\phi$ is $\approx 60~\mum$ for those modules (about 60\% of the total) that do not 
suffer from significant drift field non-uniformities.
To account for the fact that a correction for these effects was not applied
in the reconstruction of 2010 data, a systematic uncertainty 
of 300~$\mum$ was assumed for SDD points.
Along $z$, the estimated precision of the alignment 
is 50~$\mum$ for SPD and SDD and a few hundred $\mum$ for SSD.
These values are added in quadrature to the uncertainty on
the reconstructed ITS hits in the track reconstruction algorithm.
In the detector simulation, to account in an effective way for
the residual misalignment, the ITS modules are randomly displaced with respect 
to their ideal positions according to the estimated precision of the alignment.

At larger radii ($85<r<247~\cm$), the cylindrical Time Projection 
Chamber (TPC)~\cite{TPCcalib} provides track reconstruction with up to 159 three-dimensional 
space points per track, as well as particle identification via the measurement of the specific 
energy deposit $\dEdx$.
The TPC has an active
length of 500~cm along the $z$ direction and its 90~m$^3$ gas volume is 
filled with a mixture composed of 
Ne (85.7\%), CO$_2$ (9.5\%), and N$_2$ (4.8\%).
Its position resolution is 1100--1250~$\mu$m along the $z$ axis 
(corresponding to the drift direction) and 800--1100~$\mu$m along $r\phi$.
Using cosmic-ray muons and data taken in pp collisions, the relative 
$\dEdx$ resolution was measured to be 
about 5.5\% for tracks that cross the entire detector~\cite{TPCcalib}.

The charged particle identification capability of the TPC is supplemented by the 
Time-Of-Flight detector (TOF), that is based on Multi-gap Resistive Plate 
Chambers (MRPCs) in a cylindrical configuration at radius 370--399~cm from the beam axis, 
with readout consisting of
152,928 sensitive pads of dimension $2.5\times 3.5~\rm{cm^2}$.
The TOF resolution on the particle arrival time is at present better than 
100 ps. 
The start time of the collision (event time zero) is measured by the T0 
detector, an array of Cherenkov counters 
located at $+350$~cm and $-70$~cm along the beam-line, or, for the events in 
which the T0 
signal is not present, estimated using the particle arrival times at the TOF.
The particle identification is based on the difference between the measured  
time-of-flight and its expected value, computed for each mass hypothesis from the 
track momentum and length. 
The overall resolution on this difference is about 
160~ps.
In this analysis, the time-of-flight measurement 
provides kaon/pion separation up to a momentum of about $1.5~\gev/c$.
Results from the TOF commissioning with cosmic-ray particles are reported 
in~\cite{TOFcosmics}.

The data sample used for the analysis presented here consists of
314 million minimum-bias (MB) events, corresponding to an integrated luminosity 
$L_{\rm int} =5~{\rm nb}^{-1}$, collected during the 2010 LHC run with pp
collisions at $\sqrt{s}=7~\tev$.
The  minimum-bias trigger was based on the SPD and VZERO detectors.
The VZERO detector consists of two arrays of 32 scintillators each, 
placed around the beam vacuum tube on either side of the interaction region.
The two arrays cover the pseudorapidity ranges  $-3.7 < \eta < -1.7$ 
and $2.8 < \eta < 5.1$.
Minimum-bias collisions were triggered by requiring at least one hit
in either of the VZERO counters or in the SPD ($|\eta|<2$), in coincidence 
with the arrival of 
proton bunches from both directions. 
This trigger was estimated to be sensitive to about 87\% 
of the pp inelastic cross section~\cite{crosssectionpaper}.
It was verified on Monte Carlo simulations based on the PYTHIA 6.4.21 event generator~\cite{pythia} 
(with Perugia-0 tuning~\cite{perugia0})
that the minimum-bias trigger is 100\% efficient for D mesons with $\pt>1~\gev/c$ and $|y|<0.5$.
Contamination from beam-induced background was rejected offline 
using the timing information from the VZERO and the correlation between 
the number of hits and track segments (tracklets) in the SPD detector.
The instantaneous luminosity in the ALICE experiment was limited to 
0.6--$1.2\times 10^{29}~\mathrm{cm}^{-2}\mathrm{s}^{-1}$ by displacing the beams 
in the transverse plane by 3.8 times the r.m.s of their transverse profile.
In this way, the interaction probability per bunch crossing was kept in the 
range 
0.04--0.08, with probability of collision pile-up below 4\% per triggered event. 
The luminous region was measured with high precision from the distribution of 
the interaction vertices reconstructed from the charged particles tracked 
in the ALICE central barrel detectors, yielding 
$\sigma_{x}^{\rm luminous}\approx\sigma_{y}^{\rm luminous}\approx 35$--$50~\mum$ 
in the transverse plane and $\sigma_z^{\rm luminous}\approx 4$--6~cm along the 
beam direction (the quoted ranges are due to the variations of the beam 
conditions during the data taking). Only events with 
interaction vertex in the range $|z|<10$~cm were used for the analysis.

\section{Reconstruction of D meson decays}
\label{sec:decay}
The measurement of charm production was performed
by reconstructing three decay modes for D mesons, $\DtoKpi$ (with branching ratio, BR, 
of $3.87\pm 0.05\%$), $\DtoKpipi$ (BR of $9.13\pm 0.19\%$), 
and \mbox{$\rm D^*(2010)^+\to D^0\pi^+$} (strong decay with BR of $67.7\pm 0.5\%$) with $\DtoKpi$,
together with their charge conjugates~\cite{pdg}.
The $\Dzero$
and $\Dplus$ mesons have mean proper decay lengths $c\tau\approx 123$ 
and $312~\mum$, respectively~\cite{pdg}. Their decay secondary vertices 
are therefore typically displaced by a few hundred $\mum$ from the primary vertex of 
the pp 
interaction. The analysis strategy for the extraction of the $\Dzero$ and $\Dplus$ signals
from the large combinatorial background due to uncorrelated tracks 
is based on the reconstruction 
and selection of secondary vertex topologies that have significant separation from the primary vertex.
The identification of the charged kaons in 
the TPC and TOF detectors provides additional background rejection in the 
low-momentum region.
A particle identification strategy
that has high efficiency for the D meson signal (see section~\ref{sec:corrections}) was adopted.
Finally, an invariant mass analysis was used to 
extract the signal yield. 
In the $\Dstar\to\Dzero\pi^+$ case, the decay vertex cannot be resolved from the primary
vertex. The analysis exploits topological selections on the  
$\Dzero$, together with the sharp peak in the difference between the invariant
mass of the three final state hadrons and that of the two $\Dzero$ decay 
prongs. 
Given that the mass difference 
$\Delta m=m_{\rm D^{*+}}-m_{\rm D^0}\approx 145.4~\mev/c^2$~\cite{pdg}
is only slightly larger than the charged pion mass, for low $\pt$ $\Dstar$ 
mesons the produced pion has low momentum and is referred to here as a
`soft pion'.

\subsection{Track and vertex reconstruction}
\label{sec:reconstruction}

The procedure for track reconstruction in the central barrel detectors~\cite{aliceJINST,aliceStrange900} starts with a first determination of the primary vertex position, performed
by correlating hit pairs (tracklets) in the two layers of the Silicon Pixel Detector. 
The same algorithm is used to search for vertices from pile-up collisions
starting from the tracklets that do not point to the first found vertex.
An event is rejected from the analyzed data sample if a second interaction 
vertex is found, it has at least 3 associated tracklets, and it is separated 
from the first one by more than 8 mm.
The remaining undetected pile-up is negligible for the analysis described
in this paper.
Track seeds are built using this first estimate of the interaction vertex position
together with pairs of reconstructed space points in adjacent readout 
pad rows of the TPC.
Tracks are then projected inward in the radial direction
using a Kalman filter algorithm, incorporating space points in the TPC and then hits in the six layers
of the ITS (referred to here as TPC+ITS tracks).
The ITS hits not associated to TPC+ITS tracks by this procedure are then 
used to search for ITS-only tracks, 
most importantly to recover pions with $80<\pt<200~\mev/c$ 
that have very low reconstruction 
efficiency in the TPC, for geometrical reasons. ITS-only tracks are found
using a hit grouping algorithm that projects angular windows radially outward 
from the primary vertex.
The TPC+ITS tracks are then propagated outward in order to associate signals
in the large-radius detectors that perform particle identification.
Finally, all tracks  
are re-propagated with the Kalman filter in the inward direction. The relative $\pt$ resolution 
at the primary vertex for this procedure is about $1\%~(6\%)$ at $1~\gev/c$ for TPC+ITS 
(ITS-only) tracks.
The last step of event reconstruction is the re-determination 
of the primary vertex position from the accepted tracks~\cite{noteVtx2009}.
The primary vertex coordinates and covariance matrix are obtained via an 
analytic $\chi^2$ minimization method applied to the tracks approximated
by straight lines after propagation to their common point of closest 
approach.
The algorithm is then repeated excluding the tracks with distance to the 
primary vertex, normalized to its estimated uncertainty, larger than 3, which 
are incompatible with being produced by primary particles. 
The vertex fit is constrained in the transverse plane 
using the information on the position and spread 
$\sigma_{x,y}^{\rm luminous}$
of the luminous region. The latter is determined 
from the distribution of primary vertices averaged over the run
and is tabulated as a function of time during the full data-taking period.
The position resolution of the primary vertex reconstructed from tracks
depends on the particle multiplicity.
It was measured to be
$\sigma_z\,(\mum)\approx 430/N_{\rm tracklets}^{0.7}$ in the longitudinal 
direction and 
$\sigma_{x,y}\,(\mum)\approx {\rm min}(\sigma_{x,y}^{\rm luminous},600/N_{\rm tracklets}^{0.9})$
in the transverse coordinates by fitting its dependence on the number of 
tracklets in the SPD  ($N_{\rm tracklets}$), which corresponds to about twice 
the multiplicity of primary charged particles per unit of rapidity. 
Thus, for the average
 luminous region spread $\sigma_{x,y}^{\rm luminous}\approx 40~\mum$, the transverse 
position of the vertex has a resolution that ranges from $40~\mum$ in 
low-multiplicity events (i.e. below 10 charged particles per unit of rapidity) 
to about 10~$\mum$ in events with a multiplicity of about 40.

\begin{figure}[!t]
  \begin{center}
  \includegraphics[width=0.49\textwidth]{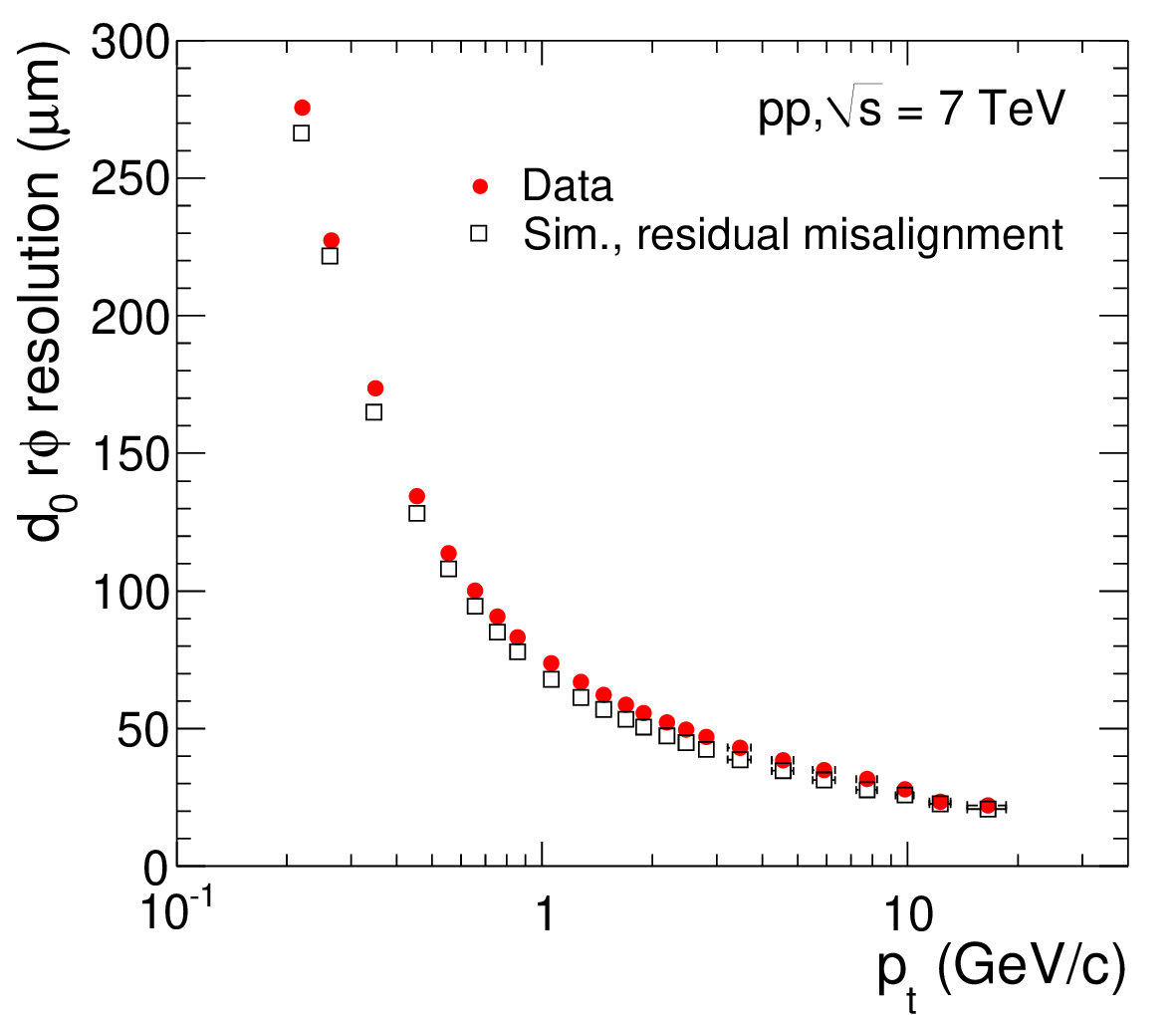}
  \caption{Track impact parameter ($d_0$) resolution in the transverse plane 
($r\phi$ direction)
              as a function of $\pt$ comparing data and simulation. This resolution includes the uncertainty in the primary vertex position, which is reconstructed excluding the track being probed.}
\label{fig:d0res}
\end{center}
\end{figure}

Secondary vertices of $\Dzero$ and $\Dplus$ meson candidates were reconstructed using tracks
having $|\eta|<0.8$, $\pt>0.4~\gev/c$, at least 70 associated 
space points (out of a maximum of 159) 
and $\chi^2/{\rm ndf}<2$ in the TPC, and at least one hit in
either of the two layers of the SPD. For tracks passing this selection, the average number of
hits in the six ITS layers is 4.5--4.7, depending on the data taking period. 
This quantity is influenced by the fraction of inactive channels 
and its distribution is 
well reproduced in Monte Carlo simulations. 
For the $\Dstar$ soft pion, all TPC+ITS and ITS-only tracks with at least 4 hits in the ITS, 
including at least one in the SPD, and $\pt>80~\mev/c$
were considered.
Figure~\ref{fig:d0res} shows the transverse momentum dependence of
the transverse ($r\phi$) 
impact parameter ($d_0$) resolution achieved with the present ITS alignment precision
for tracks that satisfy the TPC and ITS selection criteria, for data and 
Monte Carlo simulations. 
The simulations utilize GEANT3~\cite{geant3} and incorporate
a detailed description of the detector material, geometry and response. 
Proton--proton collisions were simulated using the PYTHIA 6.4.21 event generator~\cite{pythia}
with Perugia-0 tuning~\cite{perugia0}. 
 The impact parameter resolution was estimated 
by fitting the inclusive distribution of $d_0$ with respect to the 
event primary vertex, in intervals of $\pt$.
The fit function is the sum of a Gaussian, that accounts for the 
component due to prompt particles produced at the primary vertex, and two 
exponential functions,
that account for secondary particles, mainly from weak decays of strange hadrons.
The width $\sigma$ of the Gaussian provides an estimate 
of the $d_0$ resolution, which includes the 
resolution of the track parameters 
and the primary vertex position. 
In order to obtain an unbiased estimate of $d_0$, 
the primary vertex is recalculated excluding the track being probed.
The figure shows that the $d_0$
resolution measured in data, with
values of $75\,(20)~\mum$ at $\pt=1\,(15)~\gev/c$,
is reproduced within about 10\% by  
the Monte Carlo simulation incorporating the residual 
ITS misalignment described in section~\ref{sec:detector}.
The effect of the difference between data and simulation 
on the results of the D meson analysis is discussed in section~\ref{sec:systematics}.

\subsection{D meson selection}
\label{sec:selection}

$\Dzero$, $\Dplus$, and $\Dstar$ candidates were filtered by applying kinematical and topological cuts, and particle identification criteria. 
A fiducial acceptance cut $|y_{\rm D}|<y_{\rm fid}(\pt)$ was applied, with $y_{\rm fid}$ 
smoothly increasing from 0.5 to 0.8 in $0<\pt<5~\gev/c$ and  $y_{\rm fid}=0.8$ above
$5~\gev/c$.
For $\Dzero$ and $\Dplus$ decays, the secondary vertex was reconstructed 
with the same algorithm used to compute the primary vertex from tracks. 
The resolution on the position of $\Dzero$ and $\Dplus$ decay vertices was
estimated by Monte Carlo simulations to be of the order of $100~\mum$ with little 
$\pt$-dependence for $\pt>1~\gev/c$~\cite{noteVtx2009}.
For the $\Dzero$ and $\Dplus$ selection, the primary vertex was
recalculated for each D candidate, excluding the decay 
tracks.

The cut variables for the three mesons are described in the 
following.
The actual cut values are $\pt$ dependent and were tuned 
to optimize the statistical significance of the signal, resulting
in a selection efficiency that increases with increasing $\pt$. 
The cut values applied for D mesons at low $\pt$ are reported for 
reference in the next paragraphs.

For $\Dzero$ mesons, the two decay candidate tracks were further selected with $\pt>0.7~\gev/c$ 
($\pt>0.4~\gev/c$ for $1<\pt^{\Dzero}<2~\gev/c$) and 
$r\phi$ impact parameter significance $|d_0|/\sigma_{d_0}>0.5$.
Secondary vertices were required to have a minimum displacement of $100~\mum$ from the primary vertex
and a maximum distance of closest approach between the two tracks of $300~\mum$.
The cut $|\cos\theta^*|<0.8$, where $\theta^*$ is the angle between the kaon 
momentum in the $\Dzero$ rest frame and the boost direction, was applied to 
reduce the contamination of background candidates that do not represent real 
two-body decays and typically have large values of $|\cos\theta^*|$.
Well-displaced $\DtoKpi$ topologies are characterized by large and opposite-sign values
of the decay track $r\phi$ impact parameters ($d_0^{\pi}$ and $d_0^{\rm K}$) and good pointing of the reconstructed 
$\Dzero$ momentum to the primary vertex, i.e. a small value of the pointing angle $\theta_{\rm pointing}$
between the momentum and flight line.
Due to the strong correlation of these two features in the signal, the requirement 
$d_0^{\pi}\times d_0^{\rm K}< - (120~\mum)^2$ and $\cos\theta_{\rm pointing}> 0.8$
was found to be effective in increasing the signal-to-background ratio.

The $\Dplus$ selection is based on a similar strategy to that 
for $\Dzero$ mesons.
A looser cut on the $\pt$ of decay tracks, 0.4~GeV/$c$, was 
applied, due to the lower average momentum of the products of a three-body 
decay.
The candidate triplets were selected on the basis of the sum of the distances of the
decay tracks to the reconstructed decay vertex, the decay length, the cosine of the
pointing angle, and the sum of the squares of the $r\phi$ impact parameters 
of the three tracks.
Typical cut values for low-$\pt$ candidates are: decay length larger than 
$800~\mum$, $\cos\theta_{\rm pointing}>0.95$, and sum of the squares of the three
decay tracks impact parameters
$\Sigma\,d_0^2>(750~\mum)^2$.
The topological selection cuts are, in general, tighter than for the $\Dzero$ case
due to the larger $c\tau$ of the $\Dplus$ meson, resulting in a better 
separation
between primary and secondary vertices, and to the higher combinatorial
background in the three-particle final state. 

\begin{figure}[!t]
  \begin{center}
  \includegraphics[width=.99\textwidth]{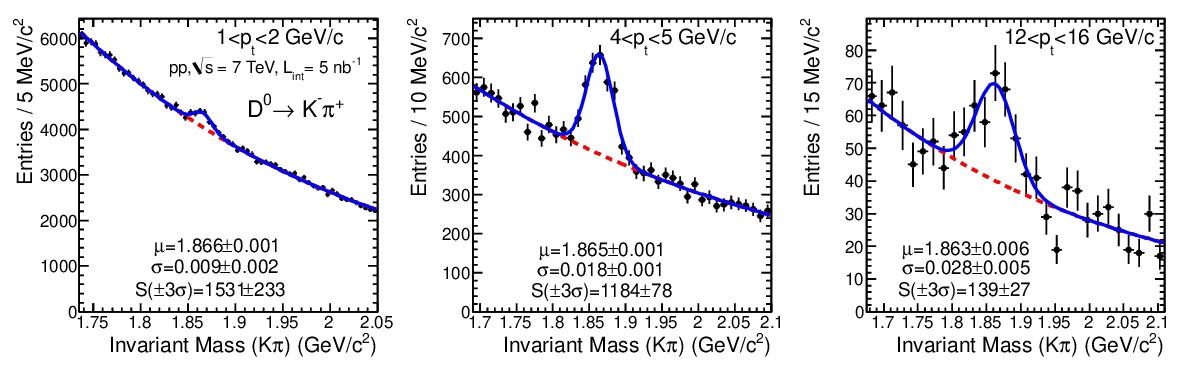}
  \includegraphics[width=.99\textwidth]{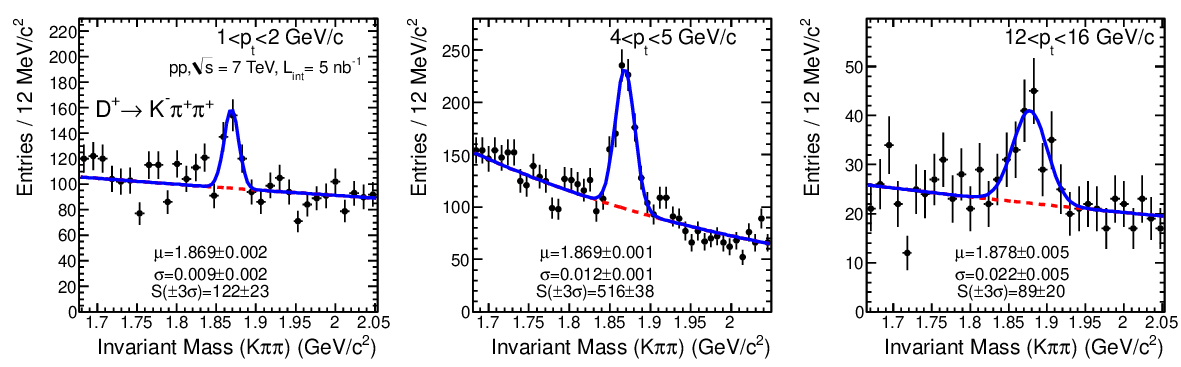}
  \includegraphics[width=.99\textwidth]{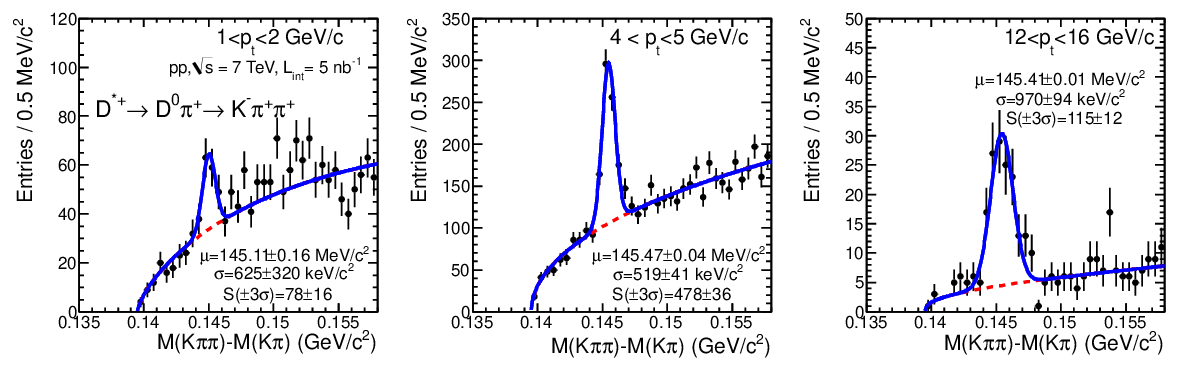}
  \caption{Invariant mass distributions for $\Dzero$ (top) and $\Dplus$ (middle) candidates, 
       and mass difference distribution for $\Dstar$ candidates (bottom), for 
         three $\pt$ intervals. The curves show the fit functions as described in the text. The values of mean ($\mu$) and width ($\sigma$) of the signal peak
are reported (for $\Dzero$ and $\Dplus$ they are expressed in~$\gev/c^2$).
}
  \label{fig:invmass}
  \end{center}
\end{figure}

The $\Dstar$ candidates were filtered by applying kinematical selections on the 
final decay products and cuts on the topology of the $\Dzero$ decay. 
The single track minimum transverse momentum was set to $0.4~\gev/c$ for the 
$\Dzero$ decay tracks and $80~\mev/c$ for the soft pion track.
The variables used to select the topology of the  $\Dzero$ decay are the same 
as for the $\Dzero$ analysis described above.
However, a selection with higher efficiency could be applied in this case, because
the background in the region around $\Delta m\approx 145~\mev/c^2$, 
which is close to the phase space boundary, is much lower than that 
around the $\Dzero$ mass.
In particular, for $\Dstar$ candidates with $\pt>6~\gev/c$, the topological 
cuts could be opened so as to select 
about 90\% of the signal passing single track cuts.

The particle identification selection used the specific energy deposit
and the time-of-flight from the TPC and TOF detectors, respectively. 
In order to assign the kaon or pion mass to the decay tracks,
compatibility cuts were applied to 
the difference between the measured and expected signals. 
For both $\dEdx$ and time-of-flight, a {\bf $3\,\sigma$} compatibility cut was 
used. 
Tracks without a TOF signal were identified using only the TPC information, and 
tracks with incompatible TOF and TPC indications were treated as non-identified, but 
still used in the analysis and considered to be compatible with both a pion and a kaon.
Two-prong candidates were accepted (as $\rm D^{0}$, $\rm \overline{D}^{0}$, or both) or rejected,
according to the compatibility with the $\rm K^\mp\pi^\pm$ final state.
For $\Dzero$ candidates used in $\Dstar$ reconstruction, compatibility with the appropriate
 final state, 
consistent with the soft pion charge, was required. Particle identification was not applied 
to the soft pion tracks.
In the case of the $\Dplus \to \rm K^- \pi^+ \pi^+$ decay, the particle with the opposite charge sign
with respect to the D meson is a kaon. Hence, the triplets were rejected 
if the opposite sign track was not compatible 
with the kaon hypothesis, or
at least one of the two same sign tracks was not compatible with the pion 
hypothesis.
For all three D meson species, a comparison of the invariant mass 
distributions obtained without and with particle identification
shows that this selection reduces the combinatorial background by a factor 2--3 
in the low $\pt$ region, while preserving close to 100\% of the D meson signal.

The raw signal yields were extracted in the intervals of $\pt$ listed in Table~\ref{tab:yields},
by a fit to the invariant mass distributions (or mass difference for the $\Dstar$), as shown in
Fig.~\ref{fig:invmass} for three selected $\pt$ intervals.
For $\Dzero$ and $\Dplus$ mesons,
the fitting function consists of a Gaussian describing 
the signal and an exponential term for the background.
In the $\Dzero$ case, the contribution of signal candidates with wrong mass 
assignment to the final state hadrons is also present in the invariant mass 
spectrum. 
It was verified on Monte Carlo simulations that this does not bias the 
extracted signal yield, because the invariant mass distribution
of these candidates is wide enough to be accounted for by the background 
function.
Moreover, at low $\pt$, the particle identification selection strongly suppresses this contribution. 
For $\Dstar$ mesons, the mass difference $\Delta m$ distribution is fitted using a 
function that consists of a Gaussian describing 
the signal and the term  
$a\,\sqrt{\Delta m-m_\pi}\cdot {\rm e}^{b(\Delta m-m_\pi)}$ for the background~\cite{charmcdf}, 
where $m_{\pi}$ is the charged pion mass, and $a$ and $b$ are free parameters. 
For all three D meson species, the mean of the Gaussian is compatible
with the PDG value~\cite{pdg} within errors,
and its width  
is well reproduced in the simulation. 
The extracted D meson raw yields are reported in Table~\ref{tab:yields}.

\begin{table}[!t]
\caption{Measured raw yields for $\Dzero$, $\Dplus$, and 
$\Dstar$ mesons, and their anti-particles, in a minimum-bias pp sample corresponding to $5~{\rm nb}^{-1}$ 
at $\sqrts=7~$TeV, 
in transverse momentum intervals. The systematic uncertainty estimation is described in 
section~\ref{sec:systematics}.}
\centering
\begin{tabular}{|c|c|c|c|} 
\hline 
$\pt$ interval & \multicolumn{3}{c|}{$N~\pm {\rm stat.} \pm {\rm syst.}$}\\
 (GeV/$c$) &  $\Dzero+\overline{\rm D}^0$ & $\Dplus+\rm D^-$ & $\Dstar+\rm D^{*-}$\\
\hline
\phantom{0}1--2\phantom{0} & $1531\pm 233\pm 340$ & $122\pm 23\pm \phantom{0}30 $ & $\phantom{0}78\pm 16\pm \phantom{0}8$\\
\phantom{0}2--3\phantom{0} & $1978\pm 168\pm 190$ & $390\pm 57\pm \phantom{0}97$ & $244\pm 26\pm 10$\\
\phantom{0}3--4\phantom{0} & $1950\pm 129\pm \phantom{0}75$ & $405\pm 40\pm 101$ & $363\pm 29\pm 11$\\
\phantom{0}4--5\phantom{0} & $1184\pm \phantom{0}78\pm \phantom{0}40$ & $516\pm 38\pm \phantom{0}46$ & $478\pm 36\pm 14$\\
\phantom{0}5--6\phantom{0} & \phantom{0}$623\pm \phantom{0}50\pm \phantom{0}25$ & $361\pm 31\pm \phantom{0}33$ & $374\pm 28\pm 18$\\
\phantom{0}6--7\phantom{0} & \phantom{0}$339\pm \phantom{0}32\pm \phantom{0}13$ & $294\pm 30\pm \phantom{0}15$ & $279\pm 22\pm 14$\\
\phantom{0}7--8\phantom{0} & \phantom{0}$199\pm \phantom{0}25\pm \phantom{0}14$ & $213\pm 27\pm \phantom{0}11$ & $170\pm 19\pm \phantom{0}9$\\
\phantom{0}8--12 & \phantom{0}$427\pm \phantom{0}38\pm \phantom{0}30$ & $434\pm 30\pm \phantom{0}22$ & $408\pm 28\pm 18$\\
12--16 & \phantom{0}$139\pm \phantom{0}27\pm \phantom{0}14$ & $\phantom{0}89\pm 20\pm \phantom{0}\phantom{0}9$ & $115\pm 12\pm 10$\\
16--24 & -- & $\phantom{0}{52\pm 14\pm \phantom{0}\phantom{0}5}$ & \phantom{0}$41\pm \phantom{0}6\pm \phantom{0}8$\\
\hline
\end{tabular}
\label{tab:yields}
\end{table}

\section{D meson cross sections}
\label{sec:crosssections}
\subsection{Corrections}
\label{sec:corrections}
The production cross sections of prompt charmed mesons were calculated as (e.g. for $\Dplus$):
\begin{equation}
  \label{eq:crosssectionD}
  \left.\frac{{\rm d}\sigma^{\rm D^+}}{{\rm d}\pt}\right|_{|y|<0.5}=
  \frac{1}{2}\frac{1}{\Delta y\,\Delta\pt}\frac{\left.f_{\rm prompt}(\pt)\cdot N^{\rm D^\pm~raw}(\pt)\right|_{|y|<y_{\rm fid}}}{({\rm Acc}\times\epsilon)_{\rm prompt}(\pt) \cdot{\rm BR} \cdot L_{\rm int}}\,.
\end{equation}
 $N^{\rm D^\pm~raw}(\pt)$ is the measured inclusive raw yield, obtained from the invariant mass analysis in each $\pt$ interval (of width $\Delta\pt$); 
 $f_{\rm prompt}$ is the prompt fraction of the raw yield;
 $({\rm Acc}\times\epsilon)_{\rm prompt}$ is the 
 acceptance times efficiency of prompt mesons, where $\epsilon$ accounts for 
 vertex reconstruction, track reconstruction and selection, and for D meson candidate selection
with the secondary vertex and particle identification cuts 
described in section~\ref{sec:decay}.  
$\Delta y$ ($=2\,y_{\rm fid}$) is the width of the fiducial rapidity coverage (see section~\ref{sec:selection}) and 
BR is the decay branching ratio~\cite{pdg}.
The factor $1/2$
accounts for the fact that the measured yields include particles and anti-particles while the cross 
sections are given for particles only.
The integrated luminosity was computed as $L_{\rm int}=N_{\rm pp,MB}/\sigma_{\rm pp,MB}$, 
where $N_{\rm pp,MB}$ and $\sigma_{\rm pp,MB}$ are the number and the 
cross section, respectively, of pp collisions passing the minimum-bias trigger condition
defined in section~\ref{sec:detector}. 
The	$\sigma_{\rm pp,MB}$ value, $62.5~\mb$, 
was derived from a measurement using a van der Meer scan~\cite{vdM} of
the cross section 
of collisions that give signals in both sides of the VZERO scintillator 
detector ($\sigma_{\rm VZERO\mbox{-}AND}$)~\cite{crosssectionpaper}.
The normalization factor, $\sigma_{\rm pp,VZERO\mbox{-}AND}/\sigma_{\rm pp,MB}\approx 0.87$, was found to be stable within 1\% in the data sample.
The uncertainty on $\sigma_{\rm pp,MB}$ is determined by the systematic uncertainty of 3.5\%
on $\sigma_{\rm VZERO\mbox{-}AND}$, 
which is due to the uncertainties on the beam intensities~\cite{beamintensities} 
and on the analysis procedure related to the van der Meer scan of the signal.

\begin{figure}[!t]
  \begin{center}
  \includegraphics[width=\textwidth]{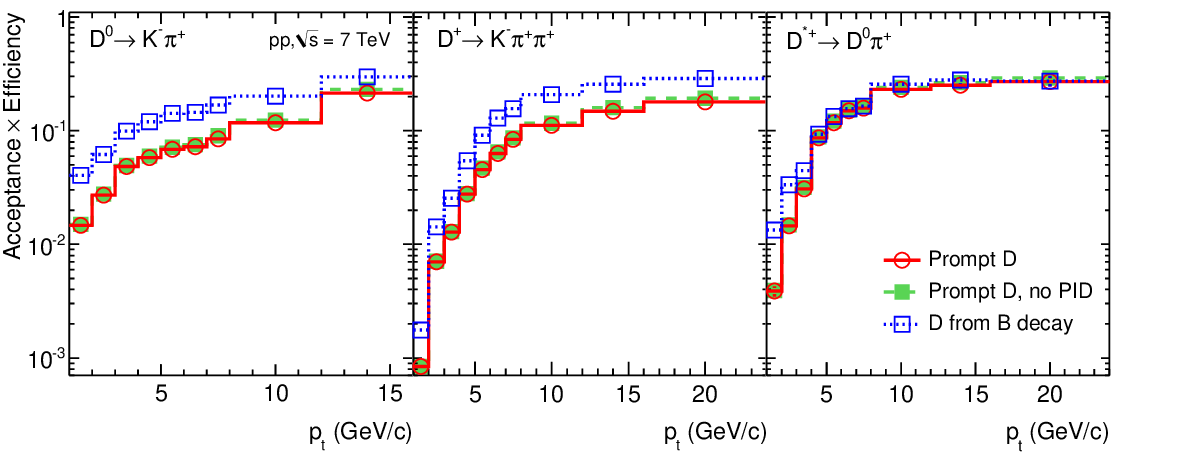}
  \caption{Acceptance $\times$ efficiency for $\Dzero$, $\Dplus$, and $\Dstar$ mesons, as a function of $\pt$ (see text for details).}
  \label{fig:efficiencies}
  \end{center}
\end{figure}

The rapidity acceptance correction, using the factor $2\,y_{\rm fid}$, with $y_{\rm fid}$ varying from 0.5 at low $\pt$ to 0.8 at high $\pt$, assumes that the 
rapidity distribution of D mesons is uniform in the range $|y|<y_{\rm fid}$. 
This assumption was checked using the PYTHIA 6.4.21 event generator~\cite{pythia}
with Perugia-0 tuning~\cite{perugia0} and the FONLL pQCD calculation~\cite{fonll,fonllydistr},
both of which generate a D meson yield that is uniform within 1\% in the range $|y|<0.8$.
The $({\rm Acc}\times\epsilon)$ correction was determined using Monte Carlo simulations
based on the GEANT3 transport code~\cite{geant3}. 
The luminous region distribution 
and the conditions of all the ALICE detectors in terms of active channels, gain, 
noise level, and alignment, and their evolution with time during the 2010 LHC run,
were included in the simulations.
Proton--proton collisions were simulated using the PYTHIA 6.4.21 event generator~\cite{pythia}
with Perugia-0 tuning~\cite{perugia0}. Only events containing D mesons were 
transported through the apparatus and reconstructed, and the efficiency was extracted 
separately for 
prompt D mesons and D mesons from B meson decays. Figure~\ref{fig:efficiencies} shows,
as a function of transverse momentum, the acceptance times efficiency $({\rm Acc\times\epsilon})$ for $\Dzero$, $\Dplus$,
and $\Dstar$ mesons with $|y|<y_{\rm fid}$.
At low $\pt$, the efficiencies are
of order 1\% or less, while for large $\pt$ the efficiencies increase and flatten at about 
10--20\% for $\Dzero$ and $\Dplus$, and 30\% for $\Dstar$.
The efficiencies without particle identification selection, shown for comparison, are the same 
as those with particle identification, 
indicating
that this selection is essentially fully efficient for the signal. 
The efficiencies for $\Dzero$ and $\Dplus$ 
mesons from B meson decays, also shown for comparison, are larger by about 
a factor of two.  This behaviour is due to the fact that 
feed-down D mesons decay further from the primary vertex,
because of the large B meson lifetime ($c\tau\approx 500~\mum$~\cite{pdg}). 
For $\Dstar$ mesons, the efficiency for the prompt and feed-down
components are the same in the $\pt$ range above $4~\gev/c$, where no strong cuts on
the separation of the $\Dzero$ decay vertex from the primary vertex are applied.

\begin{figure}[!t]
  \begin{center}
  \includegraphics[width=.49\textwidth]{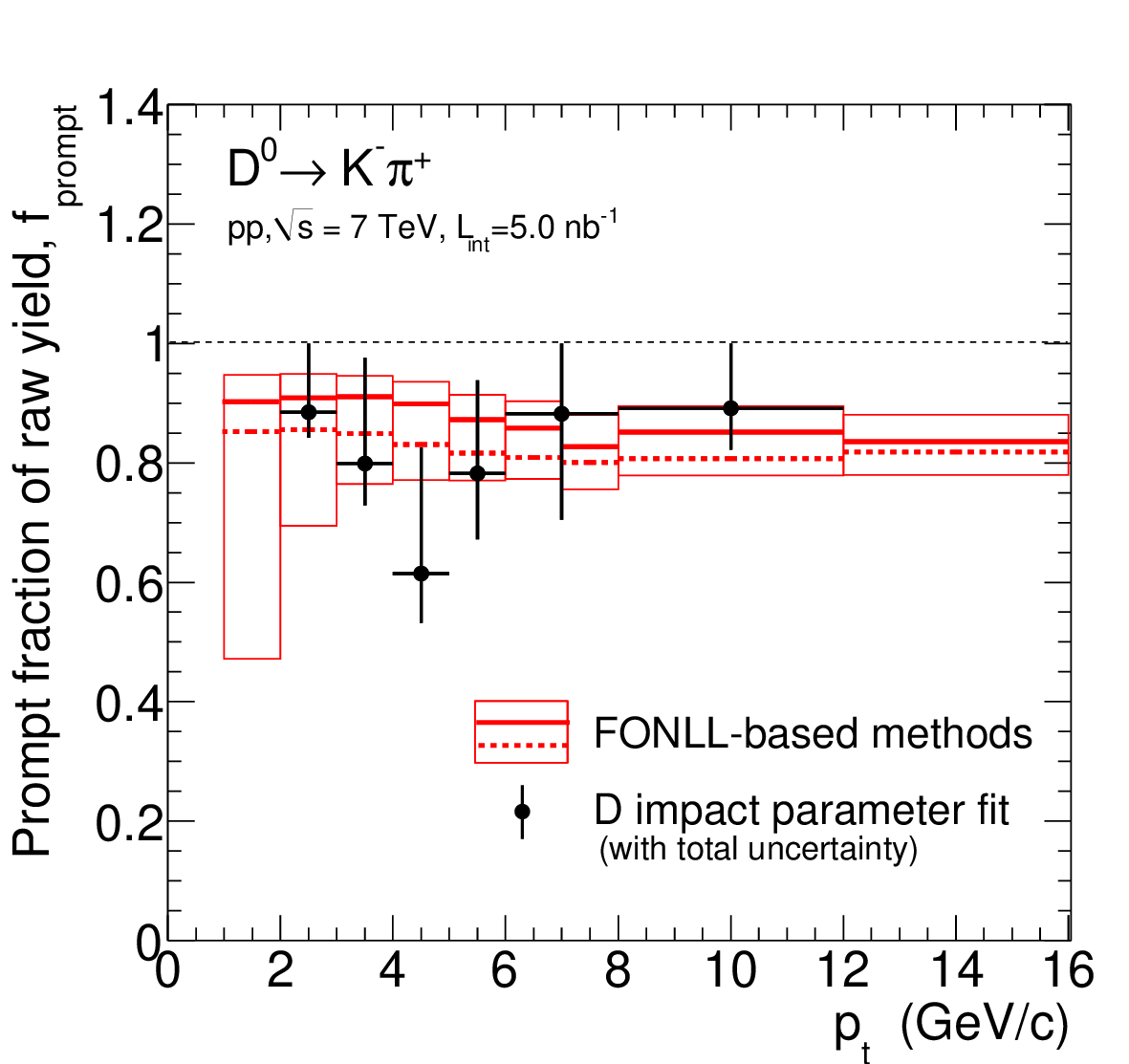}
  \includegraphics[width=.49\textwidth]{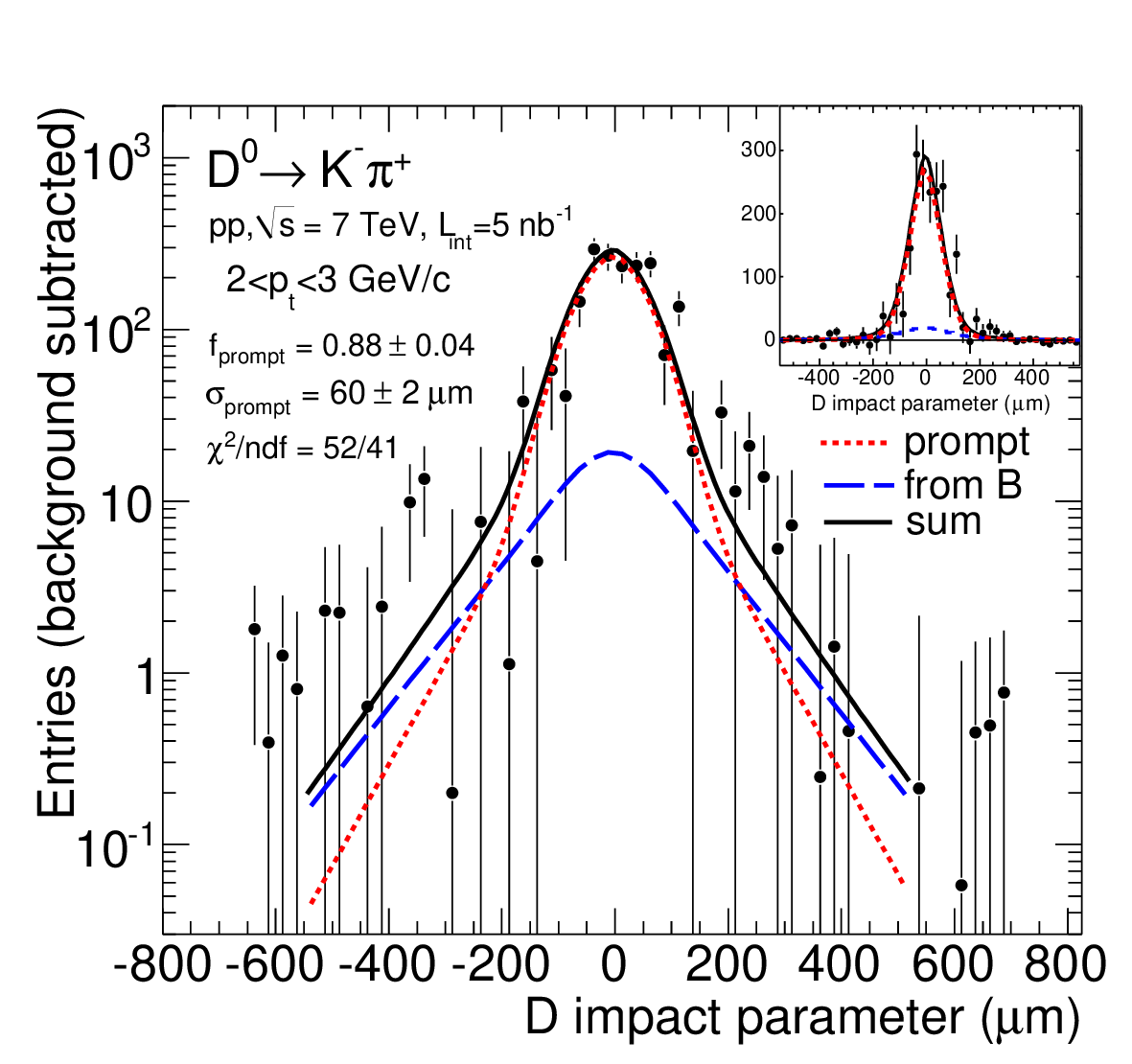}
  \caption{Left: prompt fraction $f_{\rm prompt}$ of the $\Dzero$ raw yield as a function of $\pt$, for the two FONLL-based
  methods (solid: central value, from Eq.~(\ref{eq:Nb1}); dashed: alternative method, from Eq.~(\ref{eq:fc})) and for the impact parameter fit method (circles); the boxes show the envelope
  of the uncertainty bands of the two FONLL-based methods; the error bars show the total uncertainty from the impact parameter fit, including the statistical and systematic contributions.
  Right: an example of $\Dzero$ meson impact parameter distribution in the transverse plane, 
  for $2<\pt<3~\gev/c$; 
  the distribution is background-subtracted and fitted with the two-component function for prompt
  and feed-down contributions, as described in the text; the resulting prompt fraction, impact parameter resolution for prompt mesons, and $\chi^2$/(number of degrees of freedom) of the fit are given; the inset, with linear scale, shows also the negative entries, resulting from the background subtraction.}
  \label{fig:feeddown}
  \end{center}
\end{figure}

The fraction $f_{\rm prompt}$ of D mesons coming from c quark hadronization, i.e. the 
correction factor that accounts for the feed-down from B meson decays, was evaluated using 
the B production cross section from the
FONLL pQCD calculation~\cite{fonll,fonllydistr}, which describes well beauty production 
at Tevatron~\cite{fonllBcdf} and at the LHC~\cite{lhcbBeauty,cmsJpsi}, and the $\rm B\to D$
decay kinematics from the EvtGen package~\cite{evtgen}.
The computed cross section for the feed-down component for each of the three D meson
species was 
 used, together with the Monte Carlo acceptance times efficiency $({\rm Acc}\times\epsilon)_\text{feed-down}$ for D mesons from B decays (see Fig.~\ref{fig:efficiencies}), to compute the expected feed-down contribution in the measured raw yields:
 \begin{equation}
  \label{eq:Nb1}
f_{\rm prompt}=1-(N^{\rm D^\pm~from~B~raw}/N^{\rm D^\pm~raw})
\end{equation}
with:
 \begin{equation}
  \label{eq:Nb2}
\left. N^{\rm D^\pm~from~B~raw}\right|_{|y|<y_{\rm fid}}=
  2\left.\frac{{\rm d}\sigma_{\rm FONLL}^{\rm D^+~from~B}}{{\rm d}\pt}\right|_{|y|<0.5}\cdot\Delta y\,\Delta\pt\cdot({\rm Acc}\times\epsilon)_\text{feed-down}\cdot{\rm BR}\cdot L_{\rm int}\,.
\end{equation}
The symbol of the $\pt$-dependence $(\pt)$ is omitted in the formulas, for brevity.  
The resulting prompt fraction $f_{\rm prompt}$ 
is shown in 
Fig.~\ref{fig:feeddown} (left-hand panel) by the solid horizontal lines, 
for the case of $\Dzero$ mesons. The prompt fraction ranges between 80\% and 90\%, 
depending on the $\pt$ interval, these values being determined also by the different efficiencies
for prompt and feed-down D mesons.
In order to estimate the systematic uncertainty, the perturbative uncertainty on the 
FONLL beauty production cross section was considered, as well as an 
alternative way of using the FONLL calculation. The former contribution was 
obtained by varying the b quark mass and the factorization and renormalization scales
as suggested in, e.g.,~\cite{FONLLrhic}. 
The alternative method consisted of computing the prompt fraction using the FONLL
cross sections for prompt and feed-down D mesons (with $\rm B\to D$ via EvtGen~\cite{evtgen} for the latter)
and their respective Monte Carlo 
efficiencies: 
\begin{equation}
  \label{eq:fc}
f_{\rm prompt}=
\left( 1+ \frac{({\rm Acc}\times\epsilon)_\text{feed-down} }{({\rm Acc}\times\epsilon)_{\rm prompt} } \frac{\left.\frac{{\rm d}\sigma_{\rm FONLL}^{\rm D^+~from~B}}{{\rm d}\pt}\right|_{|y|<0.5}}{\left.\frac{{\rm d}\sigma_{\rm FONLL}^{\rm D^+}}{{\rm d}\pt}\right|_{|y|<0.5}}\right)^{-1}\,.
\end{equation}
The resulting prompt fraction is shown by the dashed horizontal lines 
in Fig.~\ref{fig:feeddown} (left-hand panel). The full envelope of the uncertainty bands from the two 
methods, which is shown by the boxes in the figure, was taken as
a systematic uncertainty. The uncertainty related to the B decay kinematics was disregarded, after 
verifying that the difference resulting from using the PYTHIA~\cite{pythia} decayer instead of 
EvtGen~\cite{evtgen} is negligible with respect to
the FONLL B meson cross section uncertainty.

The prompt fraction of $\Dzero$ mesons in the reconstructed yield
was also estimated with a data-driven method based on the measured impact
parameter distribution of $\Dzero$ meson candidates in each $\pt$ interval, as done previously
by the CDF Collaboration~\cite{charmcdf}.
This method exploits the 
different shapes of the distributions of the impact parameter to the 
primary vertex of prompt and feed-down (displaced) D mesons. 
The impact parameter distribution of $\Dzero$ mesons was obtained from
the one measured for candidates with invariant mass in the range
$|m-M_{\rm D^0}|<2\,\sigma$, after subtracting the background contribution
estimated from the candidates in the side-bands (in the range $4.5\,\sigma<|m-M_{\rm D^0}|<4.5\,\sigma+100~\mev/c^2$).
The prompt fraction was estimated by fitting the resulting impact parameter 
distribution
with a two-component function.
The first component is a detector resolution term, modelled by a Gaussian and an 
exponential term, describing the impact parameter of prompt D mesons.
The second component accounts for the reconstructed impact parameter 
distribution of D from B decay, which is modelled by
a convolution of the same detector resolution term with a 
double-exponential function describing the true impact parameter of 
secondary D mesons.
The fit parameters are the width of the Gaussian and the 
fraction of prompt D mesons, that is,
the relative weight of the prompt and secondary D meson components. 
An example of such a fit is shown in Fig.~\ref{fig:feeddown} (right-hand panel)
for the $\rm D^{0}$ mesons in the transverse momentum
interval $2<\pt<3~\gev/c$. 
The prompt fraction of $\Dzero$ mesons, measured with this method for 
$2<\pt<12~\gev/c$, is shown by the
circles in Fig.~\ref{fig:feeddown} (left-hand panel) and is found to be in general 
agreement with the FONLL-based estimations.
Because of the large background for $\pt<2~\gev/c$
and the poor statistics available for $\pt>12~\gev/c$,
this method was used only as a check of the FONLL-based prompt fraction estimation,
 for the $\pt$ intervals with large signal yield.

\subsection{Systematic uncertainties}
\label{sec:systematics}
Several sources of systematic uncertainty were considered, 
including those affecting the signal extraction from the 
invariant mass spectra and all the correction factors 
applied to obtain the $\pt$-differential cross sections.
A summary of the estimated relative systematic uncertainties is given in Table~\ref{tab:Syst},
for the lowest and highest $\pt$ interval (see Table~\ref{tab:yields})
for each meson species. All the uncertainties evolve monotonously as a function of $\pt$,
except for the yield extraction and feed-down correction contributions, which are 
larger at low and high $\pt$ than at intermediate $\pt$.

\begin{table}[b!]
\caption{Summary of relative systematic uncertainties for the lowest and highest 
$\pt$ interval for each meson species.} 
\centering
\begin{tabular}{|l|cc|cc|cc|} 
\hline 
 & \multicolumn{2}{c|}{$\Dzero$} 
 & \multicolumn{2}{c|}{$\Dplus$} 
 & \multicolumn{2}{c|}{$\Dstar$}\\
 & Low $\pt$ & High $\pt$ 
 & Low $\pt$ & High $\pt$
 & Low $\pt$ & High $\pt$\\

\hline
Raw yield extraction      & 20\% & 10\% 
                      & 25\% & 10\% 
                      & 10\% & 20\% \\
Tracking efficiency   & \multicolumn{2}{c|}{{8\%}} 
                      & \multicolumn{2}{c|}{{12\%}}
                      & {13\%} & {12\%} \\
Cut efficiency        & {10\%} & {10\%}
                      & 10\% & 10\%
                      & {22\%} & {10\%}\\
PID efficiency        & {5\%} & {3\%} 
                      & {15\%} & {5\%} 
                      & {4\%} & {3\%}\\
MC $\pt$ shape        & {3\%} & {1\%} 
                      & {3\%} & {1\%}    
                      & { 3\%} & {1\%}\\
Feed-down from B      & $^{+\phantom{0}5}_{-45}\%$ & $^{+\phantom{0}8}_{-10}\%$
                  & $^{+\phantom{0}3}_{-40}\%$ & $^{+\phantom{0}8}_{-10}\%$
                  & $^{+\phantom{0}4}_{-45}\%$ & $^{+3}_{-7}\%$\\
Branching ratio       & \multicolumn{2}{c|}{1.3\%} 
                      & \multicolumn{2}{c|}{2.1\%} 
                      & \multicolumn{2}{c|}{1.5\%}\\
\hline
Normalization         & \multicolumn{6}{c|}{3.5\%}\\
\hline
\end{tabular}
\label{tab:Syst}
\end{table}

The systematic uncertainty on the yield extraction from the invariant 
mass spectrum in a given $\pt$ interval 
was determined by repeating the fit 
in a different 
mass range, and also varying the function to describe the background. 
A polynomial, instead of an exponential, was used for $\Dzero$ and $\Dplus$ mesons,
while a power law convoluted with an exponential and a polynomial was considered for $\Dstar$ mesons, 
instead of the function defined in section~\ref{sec:selection}.
A method based on bin counting (after 
subtraction of the background estimated from a fit in the mass side bands) 
was also used. The uncertainty was defined as the maximum variation in the extracted yields
 from these different methods. 

The systematic uncertainty related to the tracking efficiency includes the effects arising from 
track finding in the TPC, from track propagation  
from the TPC to the ITS, and from track quality selection.
It was estimated from the comparison of data and simulation and from the variation of the
track selection.
The resulting uncertainty is 8\% 
for the two-body decay of $\Dzero$ mesons and 12\% for the
three-body decay of $\Dplus$ mesons.
For the $\Dstar$ case, a slightly larger systematic uncertainty of 13\% was assigned in the low 
$\pt$ region (below 3~GeV/$c$), because
the soft pion often has $\pt<150~\mev/c$ and is reconstructed  
only in the ITS. The tracking efficiency in this case has a significant uncertainty 
arising from the description of hadronic interactions in the 
simulation of the detector response.

A systematic effect can arise due to residual discrepancies between data and simulation 
for the variables used to select the signal D meson candidates.
The distributions of these variables were compared for candidates passing 
loose topological cuts, i.e. essentially for background candidates, and found 
to be well described in the simulation.
The systematic effects due to residual differences between data and simulation 
were quantified by repeating the analysis with different sets of cuts.
In particular, the cut values were changed in order to vary the signal by at least $20\%$ below 
$\pt=8~\gev/c$.  
From the corresponding variation of the corrected spectra, a systematic uncertainty  
of about 10\%
was estimated for each D meson species. 
As a further cross-check, the secondary vertices in the 
simulation were reconstructed also
after a track-by-track scaling by a factor 1.08 of the impact parameter residuals with respect to their true value. This scaling is aimed at reproducing the impact parameter resolution observed in the data
(see Fig.~\ref{fig:d0res}) 
and accounts for possible residual detector misalignment effects not fully described in the simulation.
The resulting variation of the 
efficiency was found to decrease from 4\% at $\pt=1$--$2~\gev/c$ to less than $1\%$ for 
$\pt>5~\gev/c$. This effect was not included explicitly in the uncertainty estimation, 
since it is to some extent accounted for 
in the cut variation study and its magnitude is much smaller than the 10\%
uncertainty assigned to the cut efficiency corrections.

The systematic uncertainty induced by a different efficiency for particle 
identification in data and simulation was evaluated by
repeating the analysis either without applying this selection, or with a tighter 
selection ($2\,\sigma$ compatibility instead of $3\,\sigma$). The variation  
of the corrected yields obtained without PID and with these two selections (standard and 
tighter) was assigned as a systematic uncertainty.
Typical values are 3--5\%, with the exception of 15\% for $\Dplus$ in 1--$2~\gev/c$.

The accuracy of the description of the evolution of the experimental conditions with 
time was verified by analyzing separately sub-samples of data
collected during different periods and with
different orientations of the magnetic field. 
The results were found to be compatible within statistical uncertainties for all 
three meson species. 
Furthermore, the $\pt$-differential yields for each D meson measured separately
for particles and anti-particles
were found to be in agreement
within statistical uncertainties.

The effect of the shape of the simulated D mesons spectrum within our 
$\pt$ intervals was 
estimated from the relative difference in the Monte Carlo efficiencies 
obtained with the $\pt$
shapes from PYTHIA~\cite{pythia} with Perugia-0 tune~\cite{perugia0} 
and from the FONLL pQCD calculation~\cite{fonll,FONLLalice}.
These two models predict a significantly different slope at high $\pt$ 
(${\rm d} N/{\rm d}\pt\propto \pt^{-4.8}$ for FONLL and $\propto \pt^{-2.5}$ for PYTHIA
Perugia-0), 
which however results in a systematic effect on the D meson
selection efficiency of only $3\%$ for $1<\pt<2~\gev/c$, and less than $1\%$ at higher $\pt$.

The systematic uncertainty from the subtraction of feed-down D mesons from B decays
was estimated as described in the previous section. It ranges between $^{+\phantom{0}5}_{-45}\%$ 
at low $\pt$ (1--2~GeV/$c$) and $^{+\phantom{0}8}_{-10}\%$ at high $\pt$ ($>12~\gev/c$).

Finally, the results have global systematic uncertainties caused by the branching ratios (taken from~\cite{pdg}) and by the minimum-bias cross section (3.5\%).

\subsection{Results}
\label{sec:results}

\begin{figure}[!t]  
\begin{center}        
\includegraphics[width=.49\textwidth]{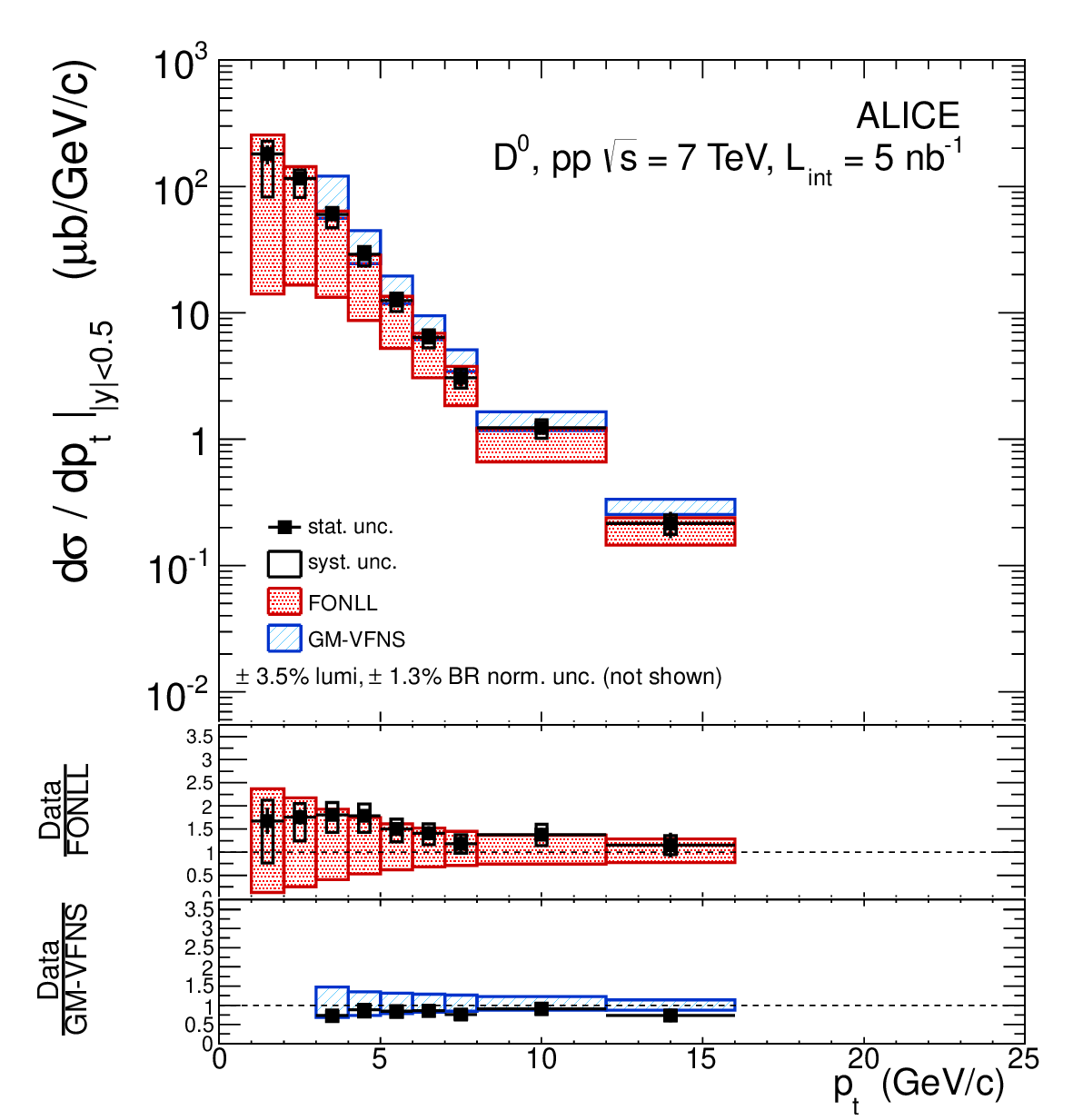}
\includegraphics[width=.49\textwidth]{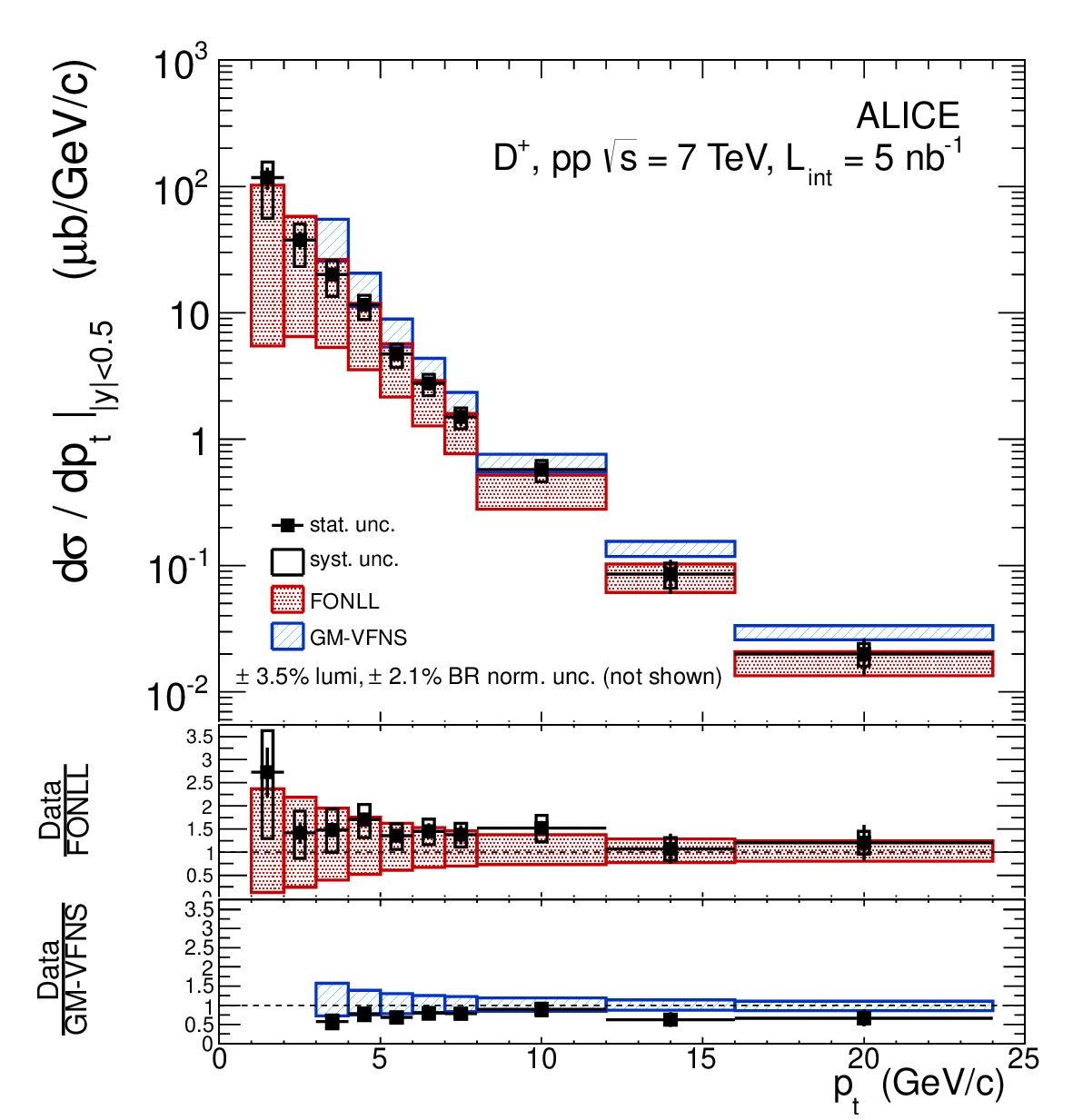}
\includegraphics[width=.49\textwidth]{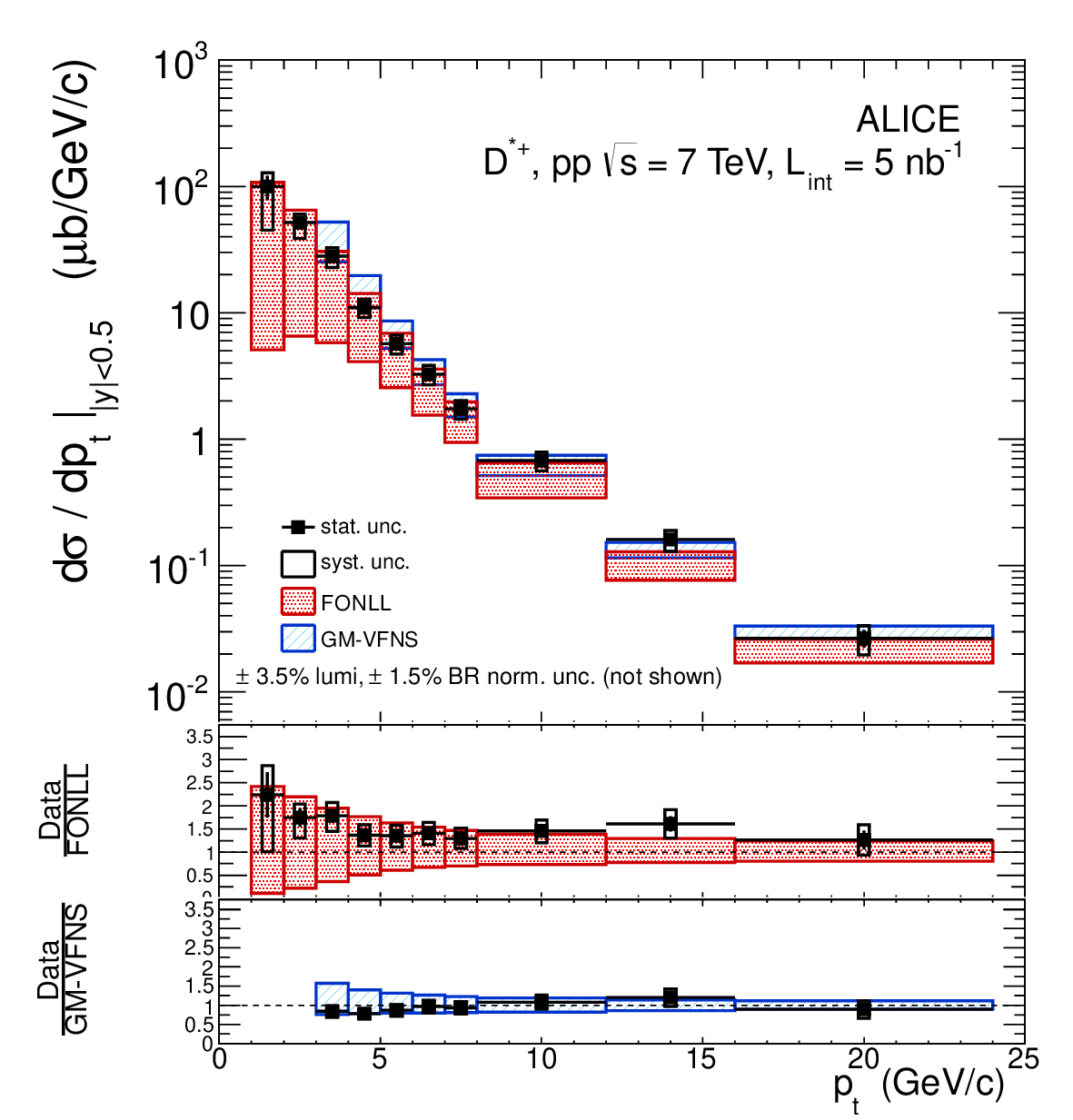}
\caption{(colour online) $\pt$-differential inclusive cross section for prompt $\Dzero$, $\Dplus$, and 
$\Dstar$ mesons in pp collisions at $\sqrts=7~$TeV compared with FONLL~\cite{fonll,FONLLalice} 
and GM-VFNS~\cite{gmvfnsDcdf,VFNSalice} theoretical predictions. The symbols are positioned 
horizontally at the centre of each $\pt$ interval. The normalization uncertainty is not shown (3.5\% from the minimum-bias cross section plus the branching ratio uncertainties, as of Table~\ref{tab:Syst}).}
\label{fig:ptdiffcs}
\end{center}
\end{figure}

The $\pt$-differential inclusive cross sections
are shown in Fig.~\ref{fig:ptdiffcs} for prompt $\Dzero$, 
$\Dplus$, and $\Dstar$ mesons.
The error bars represent the statistical uncertainties, while the systematic uncertainties
are shown as boxes around the data points. 
The numerical values for the differential 
cross sections are reported in Table~\ref{tab:ptdiffcs}, together
with their statistical and systematic uncertainties, as well as the average transverse 
momentum $\av{\pt}$ of D mesons in each $\pt$ interval.
The value of $\av{\pt}$ was obtained from the 
meson $\pt$ distribution in the considered interval, after 
subtracting the background contribution from the side bands in the invariant mass distribution.
The resulting $\av{\pt}$ values for the three D meson species are compatible within uncertainties
in all $\pt$ bins and their average is reported in the table.

\begin{table}[!b]
\caption{Production cross section in $|y|<0.5$ for prompt $\Dzero$, $\Dplus$, and 
$\Dstar$ mesons in pp collisions at $\sqrts=7~$TeV, in transverse momentum intervals.
The normalization systematic uncertainty (3.5\% from the minimum-bias cross section plus the branching ratio uncertainties, as of Table~\ref{tab:Syst}) is not included in the systematic uncertainties
reported in the table.}
\centering
\begin{tabular}{|c|c|c|c|c|} 
\hline 
$\pt$ interval & $\langle\pt\rangle$ & \multicolumn{3}{c|}{$\left.{\rm d}\sigma/{\rm d}\pt \right|_{|y|<0.5}~\pm {\rm stat.} \pm {\rm syst.}~(\mub/\gev/c)$}\\
(GeV/$c$) & (GeV/$c$) & $\Dzero$ & $\Dplus$ & $\Dstar$\\
\hline 
1--2 & $\phantom{0}1.5 \pm 0.3$ & $180\pm 30 ^{+ 48} _{- 98}$ & $117\pm 23 ^{+ 39} _{- 61}$ & $100\pm 22 ^{+ 28} _{- 55}$  \\
2--3 & $\phantom{0}2.5 \pm 0.2$ & $115\pm 11 ^{+ 20} _{- 33}$ & $37.7\pm 6.1 ^{+ 12.6} _{- 14.5}$ &  $51.8\pm 5.9 ^{+ \phantom{0}8.7} _{- 13.2}$ \\
3--4 & $\phantom{0}3.5 \pm 0.1$ & $59.7\pm 4.3 ^{+\phantom{0}8.5} _{- 12.6}$ & $20.1\pm 2.2 ^{+ 6.0} _{- 6.5}$ & $28.0\pm 2.3 ^{+ 4.6} _{- 5.2}$ \\
4--5 & $\phantom{0}4.5 \pm 0.1$ & $29.1\pm 2.1 ^{+ 4.2} _{- 5.8}$ & $11.51\pm 0.96 ^{+ 2.20} _{- 2.64}$ & $11.01\pm 0.87 ^{+ 1.82} _{- 1.88}$ \\
5--6 & $\phantom{0}5.5 \pm 0.1$ & $12.5\pm 1.1 ^{+ 1.8} _{- 2.3}$ & $4.72\pm 0.47 ^{+ 0.92} _{- 1.00}$ & $5.70\pm 0.45 ^{+ 0.97} _{- 1.00}$ \\
6--7 & $\phantom{0}6.5 \pm 0.1$ & $6.37\pm 0.70 ^{+ 0.94} _{- 1.08}$ & $2.76\pm 0.32 ^{+ 0.49} _{- 0.55}$ & $3.26\pm 0.27 ^{+ 0.55} _{- 0.57}$ \\
7--8 & $\phantom{0}7.4 \pm 0.1$ & $3.07\pm 0.47 ^{+ 0.50} _{- 0.53}$ & $1.50\pm 0.22 ^{+ 0.27} _{- 0.29}$ & $1.74\pm 0.21^{+ 0.30} _{- 0.30}$ \\
8--12 & $\phantom{0}9.4 \pm 0.3$ &  $1.23\pm 0.13 ^{+ 0.19} _{- 0.21}$ & $0.575\pm 0.056 ^{+ 0.103} _{- 0.115}$ & $0.677\pm 0.050 ^{+ 0.113} _{- 0.116}$ \\
12--16 & $13.8 \pm 0.9$ & $0.215\pm 0.050 ^{+ 0.037} _{- 0.038}$ & $0.085\pm 0.026 ^{+ 0.019} _{- 0.020}$ & $0.160\pm 0.016 ^{+ 0.030} _{- 0.031}$ \\
16--24 & $17.0_{-1.0}^{+2.0}$ & -- & $0.020\pm 0.007 ^{+ 0.004} _{- 0.004}$ & $0.027\pm 0.004 ^{+ 0.007} _{- 0.007}$ \\
\hline
\end{tabular}
\label{tab:ptdiffcs}
\end{table}

The measured D meson inclusive differential production cross sections are compared to two 
theoretical predictions, namely FONLL~\cite{fonll,FONLLalice} and 
GM-VFNS~\cite{gmvfnsDcdf,VFNSalice}. Both calculations use CTEQ6.6 parton distribution
functions (PDF)~\cite{cteq6} 
and vary the factorization and renormalization scales, $\mu_{\rm F}$ and $\mu_{\rm R}$,
independently 
in the ranges $0.5<\mu_{\rm F}/m_{\rm t}<2$, $0.5<\mu_{\rm R}/m_{\rm t}<2$, 
with the constraint $0.5<\mu_{\rm F}/\mu_{\rm R}<2$, 
where $m_{\rm t}=\sqrt{\pt^2+m_{\rm c}^2}$. The charm quark mass is
varied in FONLL within $1.3<m_{\rm c}<1.7~\gev/c^2$.
Both calculations are compatible with the measurements, within the uncertainties.
The central value of the GM-VFNS predictions lies systematically 
above the data, while that of the FONLL predictions lies below the data. 
For FONLL, this feature was observed also at $\sqrt{s}=0.2~\tev$ 
(pp)~\cite{phenixelepp,starelepp} and $1.96~\tev$ ($\rm p\overline{p}$)~\cite{charmcdf}. 
With a reach down to $\pt=1~\gev/c$, this measurement probes the gluon distribution 
in the $x$ range of a few $10^{-4}$. 
 Within the current uncertainties of the experimental measurement and of the theoretical predictions, it is not possible to draw conclusions about small-$x$ gluon saturation effects (see section~\ref{sec:intro}).

The $\pt$-integrated visible cross sections, $\sigma^{\rm vis}(\pt>1~\gev/c,\,|y|<0.5)$, for the three mesons were extrapolated down to $\pt=0$ to estimate the production cross sections per unit of rapidity ${\rm d}\sigma/{\rm d}y$ at mid-rapidity. The extrapolation factor was computed 
from the FONLL calculation~\cite{fonll,fonllydistr} as the 
ratio $({\rm d}\sigma_{\rm FONLL}/{\rm d}y)/\sigma^{\rm vis}_{\rm FONLL}$ 
and it amounts to $1.25^{+0.29}_{-0.09}$ for $\Dzero$ and $\Dplus$, and $1.21^{+0.29}_{-0.08}$ 
for $\Dstar$, for the central values of the calculation.
Its uncertainty was obtained as a quadratic sum of the uncertainties from charm mass and 
perturbative scales, varied within the aforementioned ranges\footnote{The $+0.29$ (i.e. $+23\%$) uncertainty is mainly determined by the case $\mu_F=0.5\,m_{\rm t}$, for which the PDFs are used in the region $Q\approx 0.5\,m_{\rm c}$ that is not constrained by experimental data~\cite{cteq6}. If this case is not considered, the uncertainty becomes $+13\%$ on the upper side.}, 
and from the CTEQ6.6 PDF sets~\cite{cteq6}. 
The cross sections for the three mesons are:
\[{\rm d}\sigma^{\rm D^0}/{\rm d}y= \rm 516\pm 41 (stat.) ^{+\phantom{0}69}_{-175}(syst.) \pm 18(lumi.)\pm 7(BR) ^{+120}_{-\phantom{0}37}(extr.)~\mub,\]
\[{\rm d}\sigma^{\rm D^+}/{\rm d}y= \rm 248\pm 30 (stat.) ^{+52}_{-92}(syst.) \pm 9(lumi.) \pm 5(BR)^{+57}_{-18}(extr.)~\mub,\]
\[{\rm d}\sigma^{\rm D^{*+}}/{\rm d}y= \rm 247\pm 27 (stat.) ^{+36}_{-81}(syst.) \pm 9(lumi.)\pm 4(BR) ^{+57}_{-16}(extr.)~\mub.\]

\section{Summary}
\label{sec:conclusions}
We have presented the measurement by the ALICE Collaboration 
of the inclusive differential production cross sections of prompt D mesons at 
central rapidity, in pp collisions at $\sqrt{s}=7~\tev$ within $1<\pt<24~\gev/c$.
D mesons were reconstructed in the decay channels $\DtoKpi$, $\DtoKpipi$, and $\DstartoDpi$, and their charge conjugates.  

The $\pt$-differential cross sections are reproduced within uncertainties 
by theoretical predictions based on perturbative QCD, FONLL~\cite{FONLLalice} 
and GM-VFNS~\cite{VFNSalice}. 
More in detail, the data tend to be higher than the central value of the FONLL predictions, as it was
 observed also at lower collision energies, at RHIC and at the 
Tevatron~\cite{phenixelepp,starelepp,charmcdf}. For GM-VFNS, instead, the data lie on the lower side
of the predictions, at variance with the case of Tevatron energy~\cite{charmcdf,gmvfnsDcdf}, 
indicating that the energy dependence is steeper in this model than in data.
Our results, together with existing measurements at lower energies, can contribute to a better understanding of charm production in pQCD.

Furthermore, the measurements that we have presented open new possibilities to test PDF dynamics, 
in the regime of parton fractional momentum below
$x\sim 10^{-4}$ and squared momentum transfer down to $Q^2\sim (4~\rm GeV)^2$,
 where
the onset of gluon PDF saturation effects has been conjectured~\cite{heralhc}. 
Within the uncertainties of the data and of the theoretical predictions, the framework of factorized QCD calculations provides a reasonable description of the data points down to the lowest measured transverse momentum.
However, accurate calculations incorporating saturation effects are needed in order to
draw firm conclusions on their relevance for low-momentum charm production 
at LHC energies.  

\vspace{1cm}


\newenvironment{acknowledgement}{\relax}{\relax}
\begin{acknowledgement}
\section*{Acknowledgements}
The ALICE collaboration would like to thank all its engineers and technicians for their invaluable contributions to the construction of the experiment and the CERN accelerator teams for the outstanding performance of the LHC complex.
The ALICE Collaboration would like to thank M.~Cacciari and H.~Spiesberger for providing 
the pQCD predictions that are compared to these data.
The ALICE collaboration acknowledges the following funding agencies for their support in building and
running the ALICE detector:
Department of Science and Technology, South Africa;
Calouste Gulbenkian Foundation from Lisbon and Swiss Fonds Kidagan, Armenia;
Conselho Nacional de Desenvolvimento Cient\'{\i}fico e Tecnol\'{o}gico (CNPq), Financiadora de Estudos e Projetos (FINEP),
Funda\c{c}\~{a}o de Amparo \`{a} Pesquisa do Estado de S\~{a}o Paulo (FAPESP);
National Natural Science Foundation of China (NSFC), the Chinese Ministry of Education (CMOE)
and the Ministry of Science and Technology of China (MSTC);
Ministry of Education and Youth of the Czech Republic;
Danish Natural Science Research Council, the Carlsberg Foundation and the Danish National Research Foundation;
The European Research Council under the European Community's Seventh Framework Programme;
Helsinki Institute of Physics and the Academy of Finland;
French CNRS-IN2P3, the `Region Pays de Loire', `Region Alsace', `Region Auvergne' and CEA, France;
German BMBF and the Helmholtz Association;
General Secretariat for Research and Technology, Ministry of
Development, Greece;
Hungarian OTKA and National Office for Research and Technology (NKTH);
Department of Atomic Energy and Department of Science and Technology of the Government of India;
Istituto Nazionale di Fisica Nucleare (INFN) of Italy;
MEXT Grant-in-Aid for Specially Promoted Research, Ja\-pan;
Joint Institute for Nuclear Research, Dubna;
National Research Foundation of Korea (NRF);
CONACYT, DGAPA, M\'{e}xico, ALFA-EC and the HELEN Program (High-Energy physics Latin-American--European Network);
Stichting voor Fundamenteel Onderzoek der Materie (FOM) and the Nederlandse Organisatie voor Wetenschappelijk Onderzoek (NWO), Netherlands;
Research Council of Norway (NFR);
Polish Ministry of Science and Higher Education;
National Authority for Scientific Research - NASR (Autoritatea Na\c{t}ional\u{a} pentru Cercetare \c{S}tiin\c{t}ific\u{a} - ANCS);
Federal Agency of Science of the Ministry of Education and Science of Russian Federation, International Science and
Technology Center, Russian Academy of Sciences, Russian Federal Agency of Atomic Energy, Russian Federal Agency for Science and Innovations and CERN-INTAS;
Ministry of Education of Slovakia;
CIEMAT, EELA, Ministerio de Educaci\'{o}n y Ciencia of Spain, Xunta de Galicia (Conseller\'{\i}a de Educaci\'{o}n),
CEA\-DEN, Cubaenerg\'{\i}a, Cuba, and IAEA (International Atomic Energy Agency);
Swedish Reseach Council (VR) and Knut $\&$ Alice Wallenberg Foundation (KAW);
Ukraine Ministry of Education and Science;
United Kingdom Science and Technology Facilities Council (STFC);
The United States Department of Energy, the United States National
Science Foundation, the State of Texas, and the State of Ohio.
\end{acknowledgement}
\newpage
%
%
\appendix
\section{The ALICE Collaboration}
\label{app:collab}

\begingroup
\small
\begin{flushleft}
B.~Abelev\Irefn{org1234}\And
A.~Abrahantes~Quintana\Irefn{org1197}\And
D.~Adamov\'{a}\Irefn{org1283}\And
A.M.~Adare\Irefn{org1260}\And
M.M.~Aggarwal\Irefn{org1157}\And
G.~Aglieri~Rinella\Irefn{org1192}\And
A.G.~Agocs\Irefn{org1143}\And
A.~Agostinelli\Irefn{org1132}\And
S.~Aguilar~Salazar\Irefn{org1247}\And
Z.~Ahammed\Irefn{org1225}
% ...
\renewcommand\labelenumi{\textsuperscript{\theenumi}~}
\section*{Affiliation notes}
\renewcommand\theenumi{\roman{enumi}}
\begin{Authlist}
\item \Adef{0}Deceased
\item \Adef{Dipartimento di Fisica dell'Universita, Udine, Italy}Also at: Dipartimento di Fisica dell'Universita, Udine, Italy
\item \Adef{M.V.Lomonosov Moscow State University, D.V.Skobeltsyn Institute of Nuclear Physics, Moscow, Russia}Also at: M.V.Lomonosov Moscow State University, D.V.Skobeltsyn Institute of Nuclear Physics, Moscow, Russia
\item \Adef{Institute of Nuclear Sciences, Belgrade, Serbia}Also at: "Vin\v{c}a" Institute of Nuclear Sciences, Belgrade, Serbia
\end{Authlist}
\section*{Collaboration Institutes}
\renewcommand\theenumi{\arabic{enumi}~}
\begin{Authlist}
\item \Idef{org1279}Benem\'{e}rita Universidad Aut\'{o}noma de Puebla, Puebla, Mexico
\item \Idef{org1220}Bogolyubov Institute for Theoretical Physics, Kiev, Ukraine
\item \Idef{org1262}Budker Institute for Nuclear Physics, Novosibirsk, Russia
\item \Idef{org1292}California Polytechnic State University, San Luis Obispo, California, United States
\item \Idef{org14939}Centre de Calcul de l'IN2P3, Villeurbanne, France
\item \Idef{org1197}Centro de Aplicaciones Tecnol\'{o}gicas y Desarrollo Nuclear (CEADEN), Havana, Cuba
\item \Idef{org1242}Centro de Investigaciones Energ\'{e}ticas Medioambientales y Tecnol\'{o}gicas (CIEMAT), Madrid, Spain
\item \Idef{org1244}Centro de Investigaci\'{o}n y de Estudios Avanzados (CINVESTAV), Mexico City and M\'{e}rida, Mexico
\item \Idef{org1335}Centro Fermi -- Centro Studi e Ricerche e Museo Storico della Fisica ``Enrico Fermi'', Rome, Italy
\item \Idef{org17347}Chicago State University, Chicago, United States
\item \Idef{org1118}China Institute of Atomic Energy, Beijing, China
\item \Idef{org1288}Commissariat \`{a} l'Energie Atomique, IRFU, Saclay, France
\item \Idef{org1294}Departamento de F\'{\i}sica de Part\'{\i}culas and IGFAE, Universidad de Santiago de Compostela, Santiago de Compostela, Spain
\item \Idef{org1106}Department of Physics Aligarh Muslim University, Aligarh, India
\item \Idef{org1121}Department of Physics and Technology, University of Bergen, Bergen, Norway
\item \Idef{org1162}Department of Physics, Ohio State University, Columbus, Ohio, United States
\item \Idef{org1300}Department of Physics, Sejong University, Seoul, South Korea
\item \Idef{org1268}Department of Physics, University of Oslo, Oslo, Norway
\item \Idef{org1132}Dipartimento di Fisica dell'Universit\`{a} and Sezione INFN, Bologna, Italy
\item \Idef{org1315}Dipartimento di Fisica dell'Universit\`{a} and Sezione INFN, Trieste, Italy
\item \Idef{org1145}Dipartimento di Fisica dell'Universit\`{a} and Sezione INFN, Cagliari, Italy
\item \Idef{org1270}Dipartimento di Fisica dell'Universit\`{a} and Sezione INFN, Padova, Italy
\item \Idef{org1285}Dipartimento di Fisica dell'Universit\`{a} `La Sapienza' and Sezione INFN, Rome, Italy
\item \Idef{org1154}Dipartimento di Fisica e Astronomia dell'Universit\`{a} and Sezione INFN, Catania, Italy
\item \Idef{org1290}Dipartimento di Fisica `E.R.~Caianiello' dell'Universit\`{a} and Gruppo Collegato INFN, Salerno, Italy
\item \Idef{org1312}Dipartimento di Fisica Sperimentale dell'Universit\`{a} and Sezione INFN, Turin, Italy
\item \Idef{org1103}Dipartimento di Scienze e Tecnologie Avanzate dell'Universit\`{a} del Piemonte Orientale and Gruppo Collegato INFN, Alessandria, Italy
\item \Idef{org1114}Dipartimento Interateneo di Fisica `M.~Merlin' and Sezione INFN, Bari, Italy
\item \Idef{org1237}Division of Experimental High Energy Physics, University of Lund, Lund, Sweden
\item \Idef{org1192}European Organization for Nuclear Research (CERN), Geneva, Switzerland
\item \Idef{org1227}Fachhochschule K\"{o}ln, K\"{o}ln, Germany
\item \Idef{org1122}Faculty of Engineering, Bergen University College, Bergen, Norway
\item \Idef{org1136}Faculty of Mathematics, Physics and Informatics, Comenius University, Bratislava, Slovakia
\item \Idef{org1274}Faculty of Nuclear Sciences and Physical Engineering, Czech Technical University in Prague, Prague, Czech Republic
\item \Idef{org1229}Faculty of Science, P.J.~\v{S}af\'{a}rik University, Ko\v{s}ice, Slovakia
\item \Idef{org1184}Frankfurt Institute for Advanced Studies, Johann Wolfgang Goethe-Universit\"{a}t Frankfurt, Frankfurt, Germany
\item \Idef{org1215}Gangneung-Wonju National University, Gangneung, South Korea
\item \Idef{org1212}Helsinki Institute of Physics (HIP) and University of Jyv\"{a}skyl\"{a}, Jyv\"{a}skyl\"{a}, Finland
\item \Idef{org1203}Hiroshima University, Hiroshima, Japan
\item \Idef{org1329}Hua-Zhong Normal University, Wuhan, China
\item \Idef{org1254}Indian Institute of Technology, Mumbai, India
\item \Idef{org1266}Institut de Physique Nucl\'{e}aire d'Orsay (IPNO), Universit\'{e} Paris-Sud, CNRS-IN2P3, Orsay, France
\item \Idef{org1277}Institute for High Energy Physics, Protvino, Russia
\item \Idef{org1249}Institute for Nuclear Research, Academy of Sciences, Moscow, Russia
\item \Idef{org1320}Nikhef, National Institute for Subatomic Physics and Institute for Subatomic Physics of Utrecht University, Utrecht, Netherlands
\item \Idef{org1250}Institute for Theoretical and Experimental Physics, Moscow, Russia
\item \Idef{org1230}Institute of Experimental Physics, Slovak Academy of Sciences, Ko\v{s}ice, Slovakia
\item \Idef{org1127}Institute of Physics, Bhubaneswar, India
\item \Idef{org1275}Institute of Physics, Academy of Sciences of the Czech Republic, Prague, Czech Republic
\item \Idef{org1139}Institute of Space Sciences (ISS), Bucharest, Romania
\item \Idef{org27399}Institut f\"{u}r Informatik, Johann Wolfgang Goethe-Universit\"{a}t Frankfurt, Frankfurt, Germany
\item \Idef{org1185}Institut f\"{u}r Kernphysik, Johann Wolfgang Goethe-Universit\"{a}t Frankfurt, Frankfurt, Germany
\item \Idef{org1177}Institut f\"{u}r Kernphysik, Technische Universit\"{a}t Darmstadt, Darmstadt, Germany
\item \Idef{org1256}Institut f\"{u}r Kernphysik, Westf\"{a}lische Wilhelms-Universit\"{a}t M\"{u}nster, M\"{u}nster, Germany
\item \Idef{org1246}Instituto de Ciencias Nucleares, Universidad Nacional Aut\'{o}noma de M\'{e}xico, Mexico City, Mexico
\item \Idef{org1247}Instituto de F\'{\i}sica, Universidad Nacional Aut\'{o}noma de M\'{e}xico, Mexico City, Mexico
\item \Idef{org23333}Institut of Theoretical Physics, University of Wroclaw
\item \Idef{org1308}Institut Pluridisciplinaire Hubert Curien (IPHC), Universit\'{e} de Strasbourg, CNRS-IN2P3, Strasbourg, France
\item \Idef{org1182}Joint Institute for Nuclear Research (JINR), Dubna, Russia
\item \Idef{org1143}KFKI Research Institute for Particle and Nuclear Physics, Hungarian Academy of Sciences, Budapest, Hungary
\item \Idef{org18995}Kharkiv Institute of Physics and Technology (KIPT), National Academy of Sciences of Ukraine (NASU), Kharkov, Ukraine
\item \Idef{org1199}Kirchhoff-Institut f\"{u}r Physik, Ruprecht-Karls-Universit\"{a}t Heidelberg, Heidelberg, Germany
\item \Idef{org20954}Korea Institute of Science and Technology Information
\item \Idef{org1160}Laboratoire de Physique Corpusculaire (LPC), Clermont Universit\'{e}, Universit\'{e} Blaise Pascal, CNRS--IN2P3, Clermont-Ferrand, France
\item \Idef{org1194}Laboratoire de Physique Subatomique et de Cosmologie (LPSC), Universit\'{e} Joseph Fourier, CNRS-IN2P3, Institut Polytechnique de Grenoble, Grenoble, France
\item \Idef{org1187}Laboratori Nazionali di Frascati, INFN, Frascati, Italy
\item \Idef{org1232}Laboratori Nazionali di Legnaro, INFN, Legnaro, Italy
\item \Idef{org1125}Lawrence Berkeley National Laboratory, Berkeley, California, United States
\item \Idef{org1234}Lawrence Livermore National Laboratory, Livermore, California, United States
\item \Idef{org1251}Moscow Engineering Physics Institute, Moscow, Russia
\item \Idef{org1140}National Institute for Physics and Nuclear Engineering, Bucharest, Romania
\item \Idef{org1165}Niels Bohr Institute, University of Copenhagen, Copenhagen, Denmark
\item \Idef{org1109}Nikhef, National Institute for Subatomic Physics, Amsterdam, Netherlands
\item \Idef{org1283}Nuclear Physics Institute, Academy of Sciences of the Czech Republic, \v{R}e\v{z} u Prahy, Czech Republic
\item \Idef{org1264}Oak Ridge National Laboratory, Oak Ridge, Tennessee, United States
\item \Idef{org1189}Petersburg Nuclear Physics Institute, Gatchina, Russia
\item \Idef{org1170}Physics Department, Creighton University, Omaha, Nebraska, United States
\item \Idef{org1157}Physics Department, Panjab University, Chandigarh, India
\item \Idef{org1112}Physics Department, University of Athens, Athens, Greece
\item \Idef{org1152}Physics Department, University of Cape Town, iThemba LABS, Cape Town, South Africa
\item \Idef{org1209}Physics Department, University of Jammu, Jammu, India
\item \Idef{org1207}Physics Department, University of Rajasthan, Jaipur, India
\item \Idef{org1200}Physikalisches Institut, Ruprecht-Karls-Universit\"{a}t Heidelberg, Heidelberg, Germany
\item \Idef{org1325}Purdue University, West Lafayette, Indiana, United States
\item \Idef{org1281}Pusan National University, Pusan, South Korea
\item \Idef{org1176}Research Division and ExtreMe Matter Institute EMMI, GSI Helmholtzzentrum f\"ur Schwerionenforschung, Darmstadt, Germany
\item \Idef{org1334}Rudjer Bo\v{s}kovi\'{c} Institute, Zagreb, Croatia
\item \Idef{org1298}Russian Federal Nuclear Center (VNIIEF), Sarov, Russia
\item \Idef{org1252}Russian Research Centre Kurchatov Institute, Moscow, Russia
\item \Idef{org1224}Saha Institute of Nuclear Physics, Kolkata, India
\item \Idef{org1130}School of Physics and Astronomy, University of Birmingham, Birmingham, United Kingdom
\item \Idef{org1338}Secci\'{o}n F\'{\i}sica, Departamento de Ciencias, Pontificia Universidad Cat\'{o}lica del Per\'{u}, Lima, Peru
\item \Idef{org1146}Sezione INFN, Cagliari, Italy
\item \Idef{org1115}Sezione INFN, Bari, Italy
\item \Idef{org1313}Sezione INFN, Turin, Italy
\item \Idef{org1133}Sezione INFN, Bologna, Italy
\item \Idef{org1155}Sezione INFN, Catania, Italy
\item \Idef{org1316}Sezione INFN, Trieste, Italy
\item \Idef{org1286}Sezione INFN, Rome, Italy
\item \Idef{org1271}Sezione INFN, Padova, Italy
\item \Idef{org1322}Soltan Institute for Nuclear Studies, Warsaw, Poland
\item \Idef{org1258}SUBATECH, Ecole des Mines de Nantes, Universit\'{e} de Nantes, CNRS-IN2P3, Nantes, France
\item \Idef{org1304}Technical University of Split FESB, Split, Croatia
\item \Idef{org1168}The Henryk Niewodniczanski Institute of Nuclear Physics, Polish Academy of Sciences, Cracow, Poland
\item \Idef{org17361}The University of Texas at Austin, Physics Department, Austin, TX, United States
\item \Idef{org1173}Universidad Aut\'{o}noma de Sinaloa, Culiac\'{a}n, Mexico
\item \Idef{org1296}Universidade de S\~{a}o Paulo (USP), S\~{a}o Paulo, Brazil
\item \Idef{org1149}Universidade Estadual de Campinas (UNICAMP), Campinas, Brazil
\item \Idef{org1239}Universit\'{e} de Lyon, Universit\'{e} Lyon 1, CNRS/IN2P3, IPN-Lyon, Villeurbanne, France
\item \Idef{org1205}University of Houston, Houston, Texas, United States
\item \Idef{org20371}University of Technology and Austrian Academy of Sciences, Vienna, Austria
\item \Idef{org1222}University of Tennessee, Knoxville, Tennessee, United States
\item \Idef{org1310}University of Tokyo, Tokyo, Japan
\item \Idef{org1318}University of Tsukuba, Tsukuba, Japan
\item \Idef{org21360}Eberhard Karls Universit\"{a}t T\"{u}bingen, T\"{u}bingen, Germany
\item \Idef{org1225}Variable Energy Cyclotron Centre, Kolkata, India
\item \Idef{org1306}V.~Fock Institute for Physics, St. Petersburg State University, St. Petersburg, Russia
\item \Idef{org1323}Warsaw University of Technology, Warsaw, Poland
\item \Idef{org1179}Wayne State University, Detroit, Michigan, United States
\item \Idef{org1260}Yale University, New Haven, Connecticut, United States
\item \Idef{org1332}Yerevan Physics Institute, Yerevan, Armenia
\item \Idef{org15649}Yildiz Technical University, Istanbul, Turkey
\item \Idef{org1301}Yonsei University, Seoul, South Korea
\item \Idef{org1327}Zentrum f\"{u}r Technologietransfer und Telekommunikation (ZTT), Fachhochschule Worms, Worms, Germany
\end{Authlist}
\endgroup


%
%
%
\end{document}